\documentclass[]{tCPH2e}

\usepackage{graphicx,color,amssymb,amsmath}
\usepackage{dcolumn}
\usepackage{bm}
\usepackage{bbm}

\def\>{\rangle}
\def\<{\langle}

\def\H{ {\cal H} }

\def\I{ \mathbbm{1} }

\def\x{\boldsymbol{x}}

\def\p{\boldsymbol{p}}

\def\a{\boldsymbol{a}}
\def\ba{\boldsymbol{\overline{a}}}

\def\b{\boldsymbol{b}}
\def\bb{\boldsymbol{\overline{b}}}

\def\c{\boldsymbol{c}}
\def\bc{\boldsymbol{\overline{c}}}

\def\d{\mathrm{d}}

\begin{document}

\author{David Jennings \affil{Controlled Quantum Dynamics Theory, Department of
Physics, Imperial College London, London SW7 2AZ, United Kingdom} Matthew Leifer\affil{Perimeter Institute for Theoretical Physics, 31 Caroline Street North, Waterloo, ON N2L 2Y5, Canada}}

\title{No Return to Classical Reality}

\maketitle

\begin{abstract}
  At a fundamental level, the classical picture of the world is dead,
  and has been dead now for almost a century.  Pinning down exactly
  which quantum phenomena are responsible for this has proved to be a
  tricky and controversial question, but a lot of progress has been
  made in the past few decades.  We now have a range of precise
  statements showing that whatever the ultimate laws of Nature are,
  they cannot be classical.  In this article, we review results on the
  fundamental phenomena of quantum theory that cannot be understood in
  classical terms. We proceed by first granting quite a broad notion
  of classicality, describe a range of quantum phenomena (such as
  randomness, discreteness, the indistinguishability of states,
  measurement-uncertainty, measurement-disturbance, complementarity,
  noncommutativity, interference, the no-cloning theorem, and the
  collapse of the wave-packet) that do fall under its liberal scope,
  and then finally describe some aspects of quantum physics that can
  never admit a classical understanding -- the intrinsically quantum
  mechanical aspects of Nature.  The most famous of these is Bell's
  theorem, but we also review two more recent results in this area.
  Firstly, Hardy's theorem shows that even a finite dimensional
  quantum system must contain an infinite amount of information, and
  secondly, the Pusey--Barrett--Rudolph theorem shows that the
  wave-function must be an objective property of an individual quantum
  system.  Besides being of foundational interest, results of this
  sort now find surprising practical applications in areas such as quantum
  information science and the simulation of quantum systems.
\end{abstract}


\section{Introduction}

We are constantly told that quantum theory has revolutionized our
understanding of the universe, and reveals a strange new world,
radically different from classical Newtonian mechanics -- cats can be
both alive and dead; particles can disappear and reappear behind the
moon; spooky action-at-a-distance causes instantaneous effects at the
other side of the universe; measuring one observable disturbs the
value of another in a strange, conspiratorial way. But one can press
the matter beyond the overused lines of newspaper articles and
pop-science accounts, and ask: which quantum phenomena unequivocally
force us to discard long-held, classical conceptions of the universe
in the same way that a constant speed of light for all local observers
forces us to discard the notion of absolute time? Which phenomena of
quantum theory are \emph{intrinsically} non-classical?

We might quickly point to things like ``wave-functions'' and ``Hilbert
space''\cite{Dirac,Peres, vonNeumann}, but these are simply technical
features of the mathematics of quantum theory, and on their own shed
no light on how the physics of quantum mechanics radically departs
from the classical realm. Indeed long ago Koopman and von Neumann
showed that classical mechanics itself can be formulated in Hilbert
space \cite{koop, vonN1, vonN2}.
  
Discretization of observables, such as atomic energy levels, does not
really present any dramatic shift in our perspectives -- one can
easily define a classical phase space that is discrete in positions
and momenta with permutations for dynamics. Another answer might be
that quantum reality is fundamentally probabilistic and that
``uncertainty is hard-wired into physics'', but again is this really
such a radical departure from classicality? The behaviour of the
particles in a hot cup of coffee is massively unpredictable, and we
have absolutely no chance of describing them in any precise sense --
should we view quantum physics as an exaggerated form of statistical
randomness, perhaps in which we now have an irreducibly poor
resolution of the complicated underlying details?

Yet another response might be that the collapse of the wave-function
allows us to instantaneously cause the state of the rest of the
universe to change entirely. However consider the following scenario:
both you and a friend (who lives on the other side of the galaxy) have
been given a box each. You are both told that one box contains a gold
coin while the other box contains a silver coin. Before you open your
box, you are completely ignorant as to what is in your particular box,
and so you can only predict that with probability $1/2$ you have gold,
and with probability $1/2$ you have silver. Moreover correlations
exist, you know that if you have a silver coin then your friend has a
gold coin and conversely if you have gold then they have silver. You
open the box, and to your delight, discover that you have gold. At
that same instant you also know that the box on the other side of the
galaxy must contain silver. An instantaneous collapse of the
probability distribution has taken place! Is quantum entanglement and
the collapse of the wave-function simply an exaggerated form of this
probabilistic updating?

The aim of this review is really twofold: firstly to show that the
above phenomena are \emph{not} the features of quantum physics that
overthrow a classical conception of the world, and secondly to
identify a range of deeper phenomena that do.

Of course, in order for us to provide meaningful answers we must
commit to some minimal notion of classicality. The simple criterion
that guides us is the following:

\begin{center}
  \fbox{\parbox{0.95\textwidth}{\emph{If a phenomenon of quantum
        physics also occurs within a classical statistical physics
        setting, perhaps with minor additional assumptions that don't
        violently clash with our everyday conceptions, then it should
        not be viewed as an intrinsically quantum mechanical
        phenomenon.}}}
\end{center}

This informal condition provides a standard for how surprised we
should be by any quantum phenomenon. The term ``everyday conceptions''
is intentionally vague at this point, and ultimately depends on what
the reader deems ``classically reasonable''. However the key point
here is that the more liberal you are with ``classically reasonable''
then the \emph{stricter} you are with what aspects of quantum theory
challenge your classical conception of the world. In what follows we
adopt a fairly generous notion of ``classicality'', or equivalently,
we adopt a high standard for what we call ``intrinsically
quantum''\footnote{Ultimately the formal definition of ``classical''
  will be that the theory is a ``local, non-contextual theory in which
  non-orthogonal pure states are represented by overlapping
  statistical distributions defined on some state space $\Lambda$''.}.
We start by first fleshing out the above notion of classicality, and
then exhibiting a range of quantum phenomena that, on their own, do
not seriously challenge our classical conceptions. After mapping out
these ``classical fragments'' of quantum theory, we then identify
those quantum phenomena that forever banish the classical realm.

\subsection{Overview}

This review covers several interrelated facets of the foundations of
quantum theory, which at times can become quite abstract and subtle.
To avoid confusion, we start with a rough outline.

The purpose of \S\ref{ClassFrag} is to show that the quantum phenomena
of measurement-disturbance, complementarity, randomness, collapse of
the wavepacket, and others, also appear in classical statistical
mechanics supplemented with minor additional assumptions.  \S\ref{toy}
contains the core concepts, and is largely
self-contained. \S\ref{Gaussian} provides a more physical model that
makes a direct connection to quantum physics. The conclusion of this
is that Gaussian quantum physics is essentially classical in nature
(see Appendix~\ref{CVS} for a definition of Gaussian quantum
physics). \S\ref{Cloning} and \S\ref{EPRsec} analyse the no-cloning
theorem, and wavepacket collapse in the Einstein-Podolosky-Rosen
experiment within this model, showing that they too are essentially
classical.

Leading on from this, \S\ref{Gibbs} and \S\ref{KS} discuss what it
means for quantum phenomena to have a classical statistical model in
general.  This sets the scene for discussing intrinisically
quantum-mechanical phenomena.  These sections are a bit more abstract
and technical, so a casual reader should just take the core message
from section \S\ref{Gibbs} that a probabilistic framework allows us to
place quantum theory in a broader context, and in doing so contrast it
with other theories such as classical theory.

\S\ref{NC} identifies three intrinisically quantum-mechanical
phenomena.  \S\ref{Bell} reviews Bell's theorem.  This is a mostly
self-contained discussion, and can be read on its own with only a
rough overview of \S\ref{ClassFrag}.  The take-home message is that
the correlations obtained from measuring entangled states force us
into a dilemma: either abandon basic notions of realism or abandon the
relativistic notion that influences cannot travel faster than
light. \S\ref{Hardy} discusses Hardy's theorem.  This requires
understanding the basic framework of \S\ref{Gibbs}, and the take-home
message is that quantum systems display seemingly contradictory
properties of being both continuous and discrete and, contrary to
traditional statements, it is the continuity which is
puzzling. Finally, \S\ref{PBR} discusses the recent
Pusey-Barrett-Rudolph theorem, which shows that the wave-function must
be an objective property of an individual system.  This requires a
little more background from \S\ref{Gibbs} as well as a basic
understanding of \S\ref{KS}.

\section{Classical Fragments of Quantum Theory}

\label{ClassFrag}

It is clear that quantum mechanics must accommodate some kind of
emergent ``classical properties'', and some kind of a ``classical
regime'', in which Newtonian mechanics is recovered as a limiting
case. However our goal here is not to describe a classical limit, but
rather to set up a sufficiently broad notion of classicality so that
anything which \emph{does not} fall under this notion must be deemed
intrinsically quantum in character.

In this section we show that \emph{classical fragments} exist within
quantum theory, i.e.\ there are a range of phenomena in quantum
physics that also appear in classical statistical physics supplemented
with assumptions that do not violently clash with our intuitions about
classical physics. However, if one tries to stretch this classical
framework across all of quantum physics, then ``classically
unreasonable'' features always appear -- quantum physics can never be
fit cleanly into a classical framework. By carving out these classical
fragments and delineating their boundaries, we can identify the
genuinely non-classical aspects of Nature.

\subsection{A toy classical universe with some odd features}

\label{toy}
  
\begin{figure}[htb]
  \begin{center}
    \includegraphics[width=5cm]{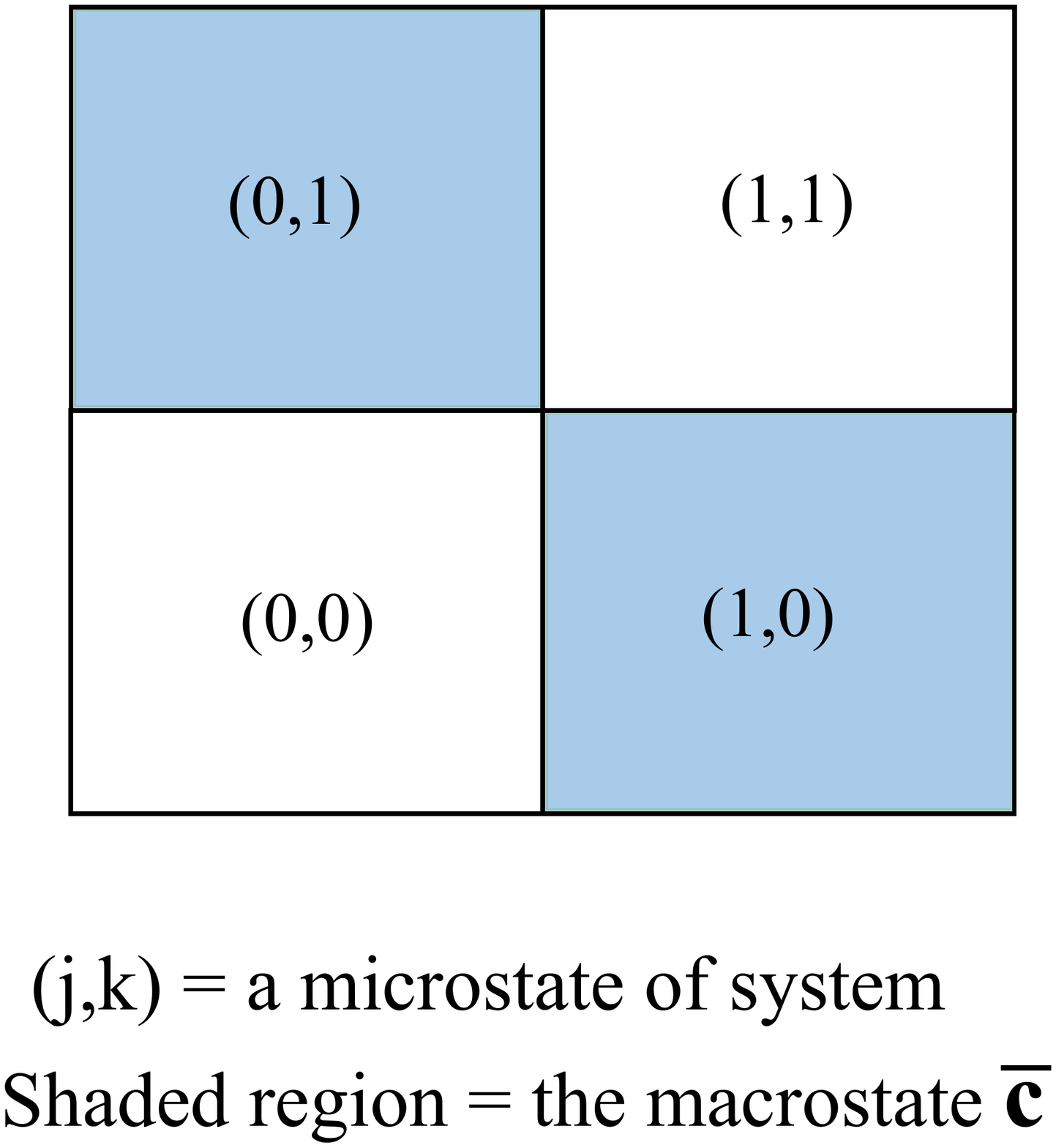}
    \includegraphics[width=7cm]{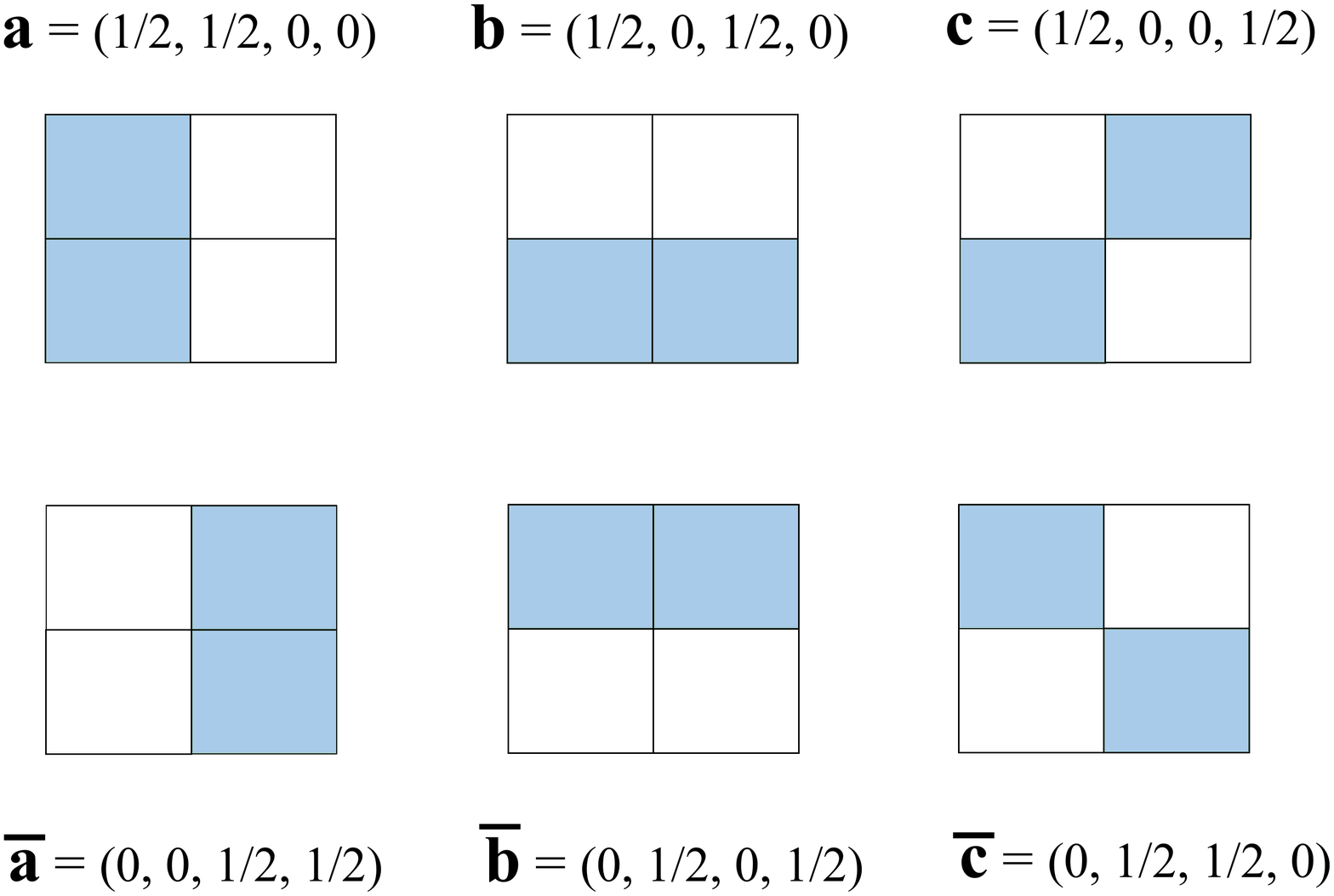}
  \end{center}
  \caption{\label{toy-states}\textbf{Extremal macrostates within the
      toy theory:} A shaded cell denotes that the particle is in that
    cell with probability $\frac{1}{2}$. Exactly two bits $j,k \in
    \{0,1\}$ are required to specify the microstate $(j,k)$ of the
    system, however if only at most one bit about $(j,k)$ is
    attainable, then there are six minimal uncertainty macrostates $\{
    \a, \b, \c, \ba, \bb, \bc\}$, within the toy theory. Every other
    macrostate is a probabilistic combination of these
    macrostates. Note that each vertical pair of macrostates have zero
    overlap, and so are perfectly distinguishable alternatives. Put
    another way, each pair defines a single answerable ``yes-no''
    question in the universe, and so explains the use of the barred
    and unbarred notation (e.g. ``a''=$a$ and ``not a''
    $=\overline{a}$.).}
\end{figure}

Let us begin with an extremely simple statistical mechanics example
due to Spekkens \cite{Spekkens-toy} that captures some of the
conceptual problems we face in identifying genuinely quantum
phenomena. Imagine a classical particle that can be in one of four
possible states. For example, we might imagine a box with four
internal cells to it, and the particle is located in one of the
cells. For simplicity we can imagine that the cells form a $2 \times
2$ grid, which allows us to label these cells via discrete coordinates
as $(0,0), (0,1), (1,0), (1,1)$. These are the exact states, or
\emph{microstates}, of the system (see Figure~\ref{toy-states}).

However, now suppose that the box is so small that all our measurement
devices are blunt, clumsy probes that only return coarse answers as to
where the particle is actually located. Instead of precise
microstates, we are forced to use \emph{statistical macrostates},
which are probability distributions $\p = (p_{(00)}, p_{(01)},
p_{(10)}, p_{(11)})$, where $p_{(jk)}$ is the probability that the
particle is in the cell $(j,k)$.

So far everything is elementary classical statistical mechanics of a
simple system, which requires exactly two bits of data, $j$ and $k$,
to specify the microstate of the particle. However, suppose we now
conjure up a new fundamental law for this toy-universe
\cite{Spekkens-toy}, and make the following assumption on our ultimate
ability to determine the microstate of the particle:
\begin{center}
  \fbox{\parbox{0.95\textwidth}{\textbf{Resolution Restriction
        (RR-condition):} \emph{It is impossible to possess more than
        a single bit of information about the microstate of the
        classical system}.}}
\end{center}
In other words, the macrostate is allowed to be fully random, $\p =
(1/4, 1/4, 1/4, 1/4)$, or to have a single sharp coordinate, $j$ say,
so that $\p=(1/2,1/2,0,0)$ is allowed. However, it cannot have any
weaker form of randomness, for example the sharp microstate
$\p=(1,0,0,0)$ is fundamentally excluded from the toy-theory.

There are six extremal macrostates in the theory, which are the six
minimal uncertainty states shown in Figure~\ref{toy-states}.

The key thing to note is that while clearly not a fundamental part of
classical mechanics, this RR-condition does not dramatically overthrow
our classical conceptions -- it simply describes a classical scenario
where we have a bound on our resolving power. However despite such
simplicity, the RR-condition has a range of surprising consequences
for the physics of this toy-universe.  For a start, it implies that
the six extremal macrostates of the theory cannot be perfectly
distinguished.  For example, if the system is prepared either in the
macrostate $\a$ or the macrostate $\b$, and you do not know which,
then there is no procedure that will tell you which is the case with
certainty.  This is because the distributions $\a$ and $\b$ overlap --
they each assign probability $1/2$ to the microstate $(0,0)$.
Therefore, if the system happens to occupy this microstate, which will
happen with probability $1/2$, then there is nothing you can possibly
do to distinguish $\a$ from $\b$.  This parallels the fact that
non-orthogonal quantum states, such as the $|\uparrow z\>$ and
$|\uparrow x\>$ states of a spin-$1/2$ particle \cite{Dirac,
  Ballentine, Peres}, cannot be perfectly distinguished. Secondly, the
RR-condition not only places restrictions on what can be measured, but
it also implies that \emph{all measurements in the toy-universe must
  uncontrollably disturb the particle}.

Since we want to respect the RR-condition in this toy-universe in the
simplest way, it is reasonable to assume that any idealized
measurement that we can perform obeys the following two conditions:
\begin{itemize}
\item \textbf{Consistency with the RR-condition:} Whenever the system
  starts in a macrostate that obeys the RR-condition, it must end up
  in a macrostate that still obeys the RR-condition after the
  measurement has been performed and we have recorded the outcome.
\item \textbf{Repeatability:} When a measurement is performed and a
  certain outcome is obtained then repeating the measurement
  immediately afterwards should yield the same outcome \cite{Peres}.
\end{itemize}

Given these two conditions, it is easy to see that there is no
measurement that reveals exactly which cell the particle is in.
Suppose such a measurement were allowed and consider, for example,
what would happen if its outcome revealed that the system was in the
$(0,0)$ cell.  Because of repeatability, the system must remain in the
$(0,0)$ cell after the measurement, since otherwise there would be
some probability of obtaining a different result upon repeating the
measurement.  However, this is incompatible with the RR-condition
because it would leave us with full information about which cell the
particle is in.  We conclude that measurements in the toy universe
must necessarily only reveal coarse-grained information about the
microstate.

Whilst we cannot do a measurement that tells us which exact cell
$(j,k)$ the particle is in, it turns out that we are allowed do a
coarse-grained measurement $A$, with outcomes $a$ and $\overline{a}$,
that returns one of two answers:
\begin{center}
  Outcome $A=a$: ``The particle is in $(0,0)$ or $(0,1)$''\\
  $\,\,$Outcome $A=\overline{a}$: ``The particle is in $(1,0)$ or
  $(1,1)$''.
\end{center}
Now consider a situation in which the system is prepared in the
macrostate $\p=\b= (1/2, 0, 1/2, 0)$, namely half the time there is a
particle in the cell $(0,0)$, the other half the time its in the cell
$(1,0)$.

\begin{figure}[htb]
  \begin{center}
    \includegraphics[height=4cm]{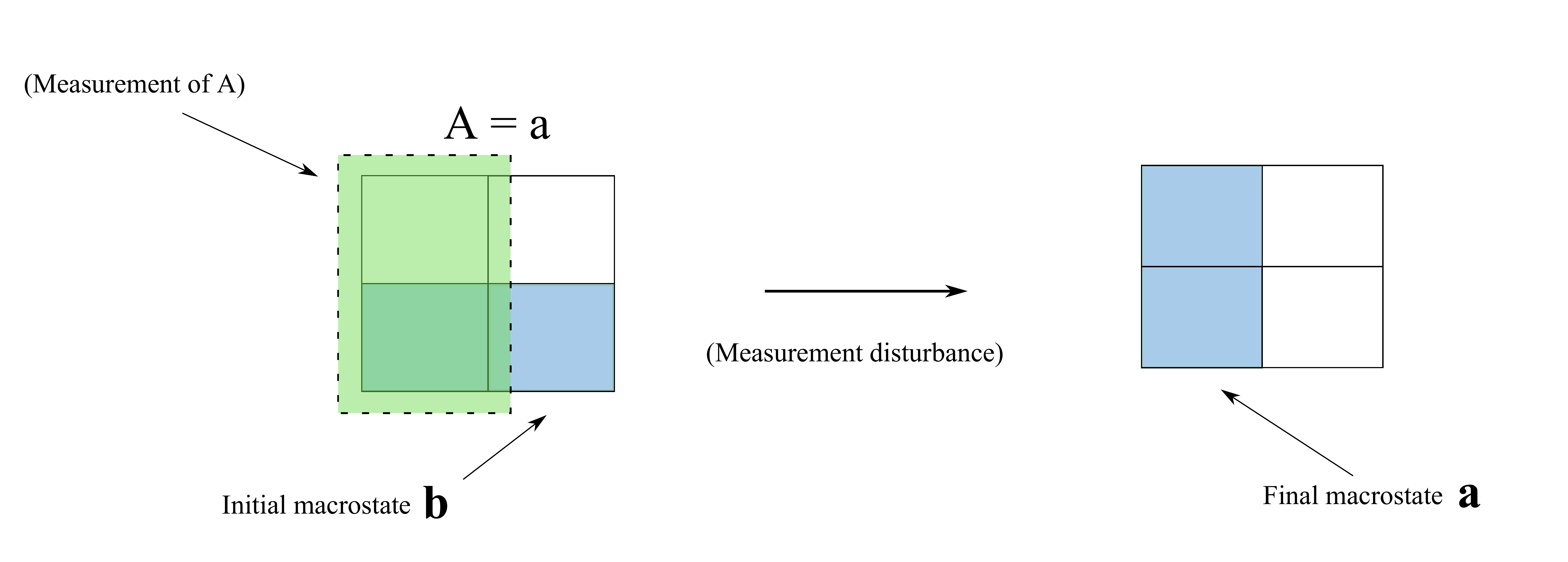}
  \end{center}
  \caption{\label{disturb}\textbf{Measurements are always disturbing
      in the toy-model:} The minimal uncertainty macrostate $\b =(1/2,
    0, 1/2,0)$, which respects the RR-condition. Suppose a measurement
    $A$ is performed on the macrostate, and randomly outputs $A=a$:
    ``top left or bottom left'' (the dashed green region), then it
    must necessarily disturb the macrostate in a probabilistic way in
    order to preserve both the RR-condition and repeatability of
    measurements.  The system is updated to the new macrostate $\a =
    (1/2,1/2,0,0)$.}
\end{figure}

If we perform the measurement $A$ then we will get the outcome $a$
half the time. Crucially, if the measurement does not cause a
disturbance to the system, then getting this outcome would allow us to
conclude that the particle must be in the $(0,0)$ cell because it had
zero probability of being in the $(0,1)$ cell to begin with.  To avoid
this violation of the RR-condition, the particle must sometimes get
kicked into another cell during the measurement procedure.  By
repeatability, the only cell it can get kicked to is $(0,1)$ because
this is the only other cell that gives the $A=a$ outcome with
certainty.  In fact, to satisfy both the RR-condition and
repeatability for any valid starting distribution, upon obtaining the
$A=a$ outcome half the time the particle must remain where it is, and
the other half the time the microstates $(0,0)$ and $(0,1)$ must be
swapped.  Thus, starting in the macrostate $\p=\b$ before the
measurement, the measurement disturbs the system and sends it to
$\p'=\a = (\frac{1}{2}, \frac{1}{2},0 ,0)$ (see
Figure~\ref{disturb}). Measurement-disturbance necessarily exists in
the physics of this toy universe. Indeed the only macrostates that are
\emph{not} disturbed by measuring $A$ are the macrostates $\a$ and
$\ba$.

\begin{figure}[htb]
  \begin{center}
    \includegraphics[width=4.5cm]{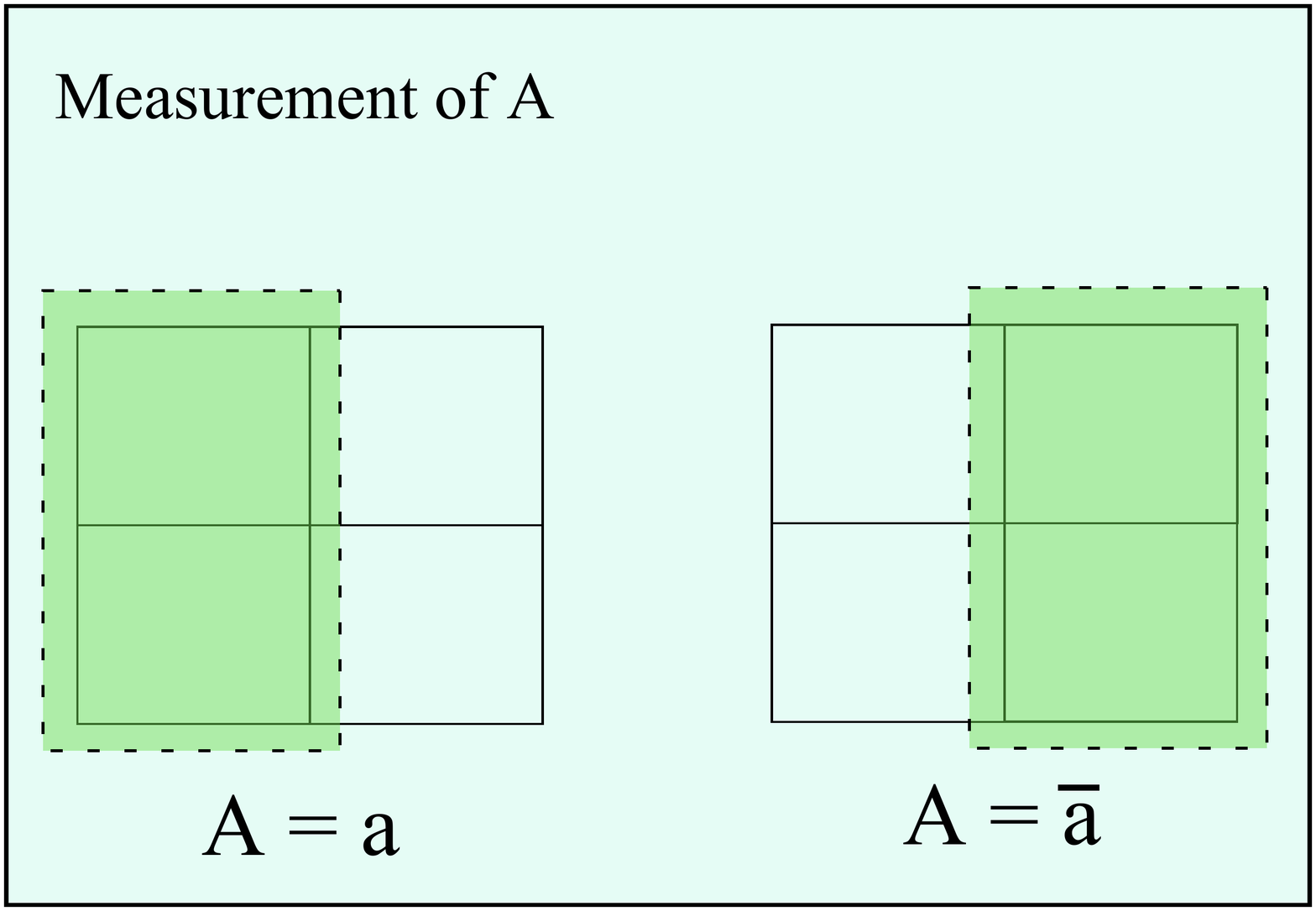}
    \includegraphics[width=4.5cm]{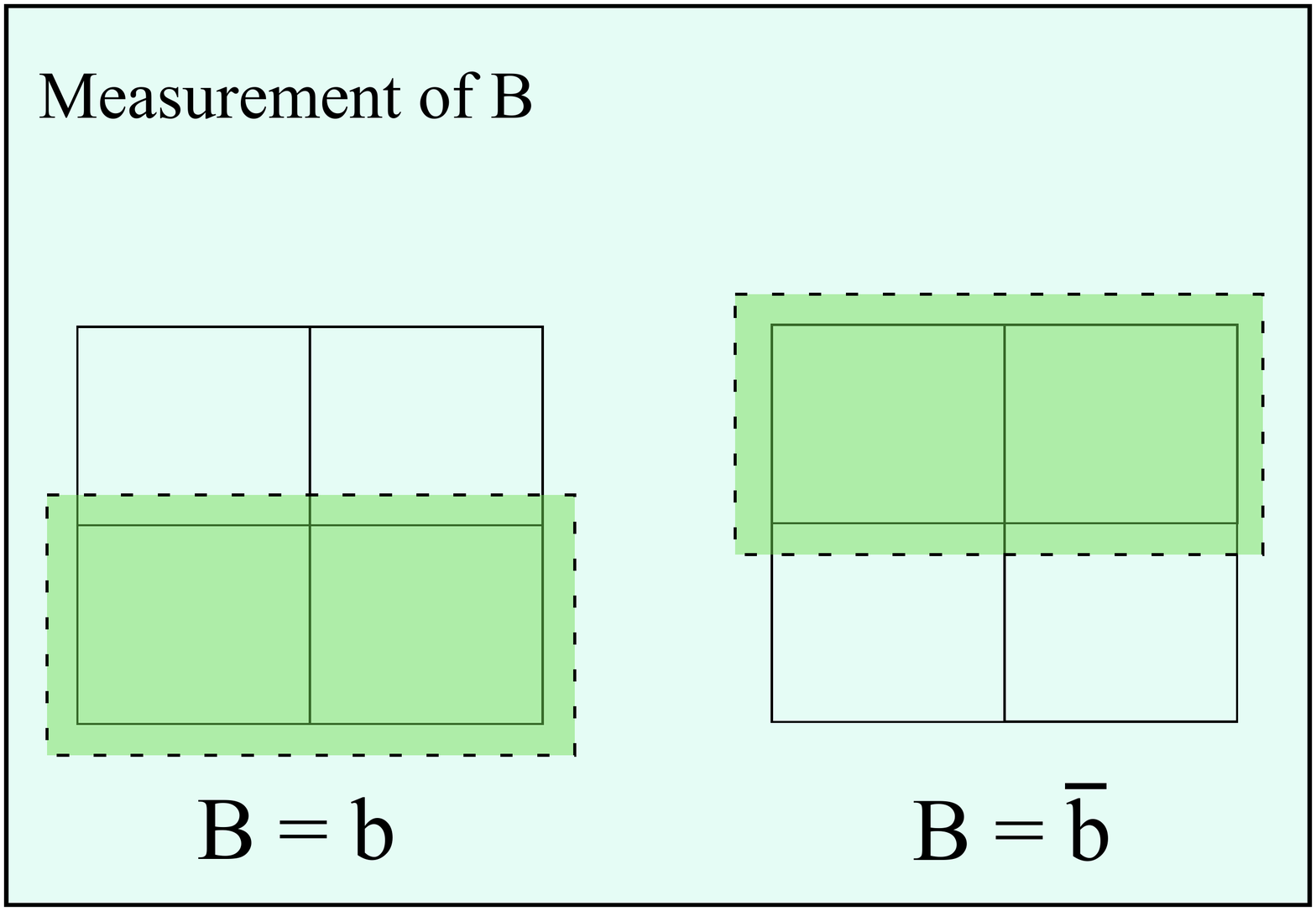}
    \includegraphics[width=4.5cm]{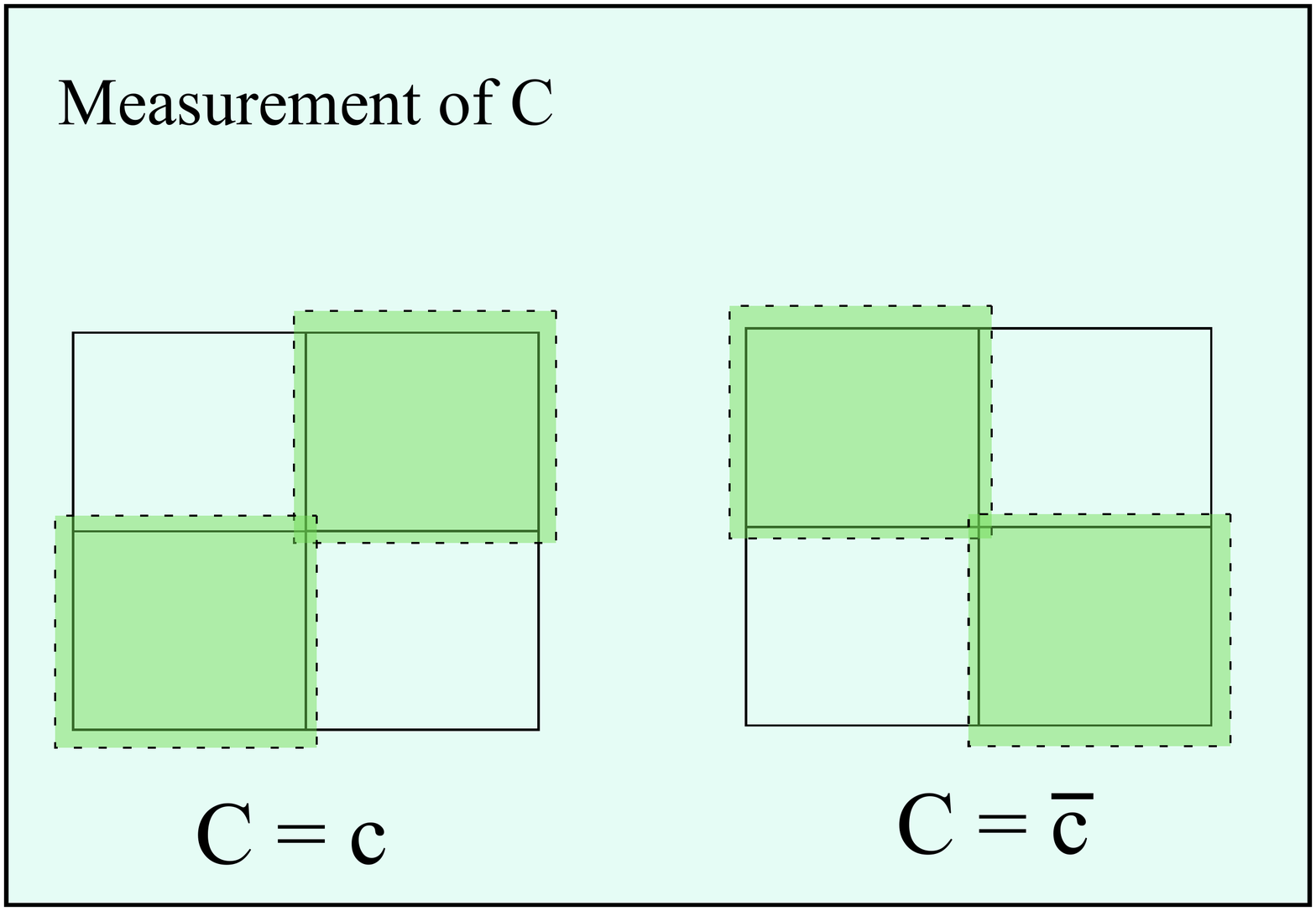}
  \end{center}
  \caption{\label{measurements}\textbf{Extremal Measurements:} The 3
    sharpest possible measurements permitted within the toy-universe
    are denoted $A, B, C$. Each measurement has two probabilistic
    outcomes, with each outcome resulting in the preparation of one of
    the 6 minimal uncertainty macrostates (within the dotted
    boxes). For example, performing the measurement $A$ on the
    completely random macrostate $(\frac{1}{4},
    \frac{1}{4},\frac{1}{4},\frac{1}{4})$ prepares the macrostate $\a$
    when we get the outcome $A=a$, and prepares the macrostate $\ba$
    when we get the outcome $A=\overline{a}$. Each of these outcomes
    occurs with probability $\frac{1}{2}$.}
\end{figure}

In addition to $A$, there are two other extremal measurements in the
toy theory: $B$ distinguishes $(0,0)$ and $(1,0)$ from $(0,1)$ and
$(1,1)$, while $C$ distinguishes $(0,0)$ and $(1,1)$ from $(0,1)$ and
$(1,0)$.  Each of these measurements necessarily induces a disturbance
of a similar type to a measurement of $A$.  The three extremal
measurements are illustrated in Figure~\ref{measurements}.

This measurement disturbance is entirely classical, however it leads
to phenomena that are familiar in quantum mechanics\footnote{In
  appendices~\ref{QT}--\ref{CVS}, we review those aspects of quantum
  mechanics that are relevant to this article.} -- for example, the
measurement of spin along the $x$-axis for a spin-1/2 particle
disturbs all quantum states except the eigenstates $|x \uparrow \>$
and $|x \downarrow \>$. As it turns out, the classical measurements
$A, B, C$ on macrostates $\{\a,\b,\c, \ba, \bb, \bc\}$ have
probabilities and disturbance patterns that perfectly mimic the three
quantum spin measurements along the $x,y,z$-directions performed on
the $6$ different eigenstates $\{|s \uparrow \> , |s \downarrow \>\}$
for $s=x,y,z$. The macrostates in the toy-model also display a form of
complementarity in terms of the 3 measurements $A, B$ and $C$. For
example, the macrostates $\a$ and $\ba$ have deterministic outcomes
for measurement $A$, however the outcomes of measurements $B$ and $C$
are fully random on these macrostates (see \cite{Peres} or
\cite{Ballentine} for careful discussions of complementarity in
quantum theory).

The disturbance induced by measurements also implies that the results
obtained in a sequence of measurements depend on the order in which
the measurements are made.  Such a dependence is often taken to be a
signature of non-commutativity in quantum mechanics, so in this sense
we can say that the classical measurements in the toy model are non-commutative.
For example, if we make two $A$ measurements in a row, one immediately
after the other, then, by repeatability, we will get the same result
both times.  However, if $B$ is measured between the two $A$
measurements, then the disturbance it induces will cause the outcome
of the second $A$ measurement to be totally random, and uncorrelated
with the first $A$ measurement.

It is vital to emphasize that this example is most definitely
\emph{not} trying to show that quantum theory is actually some funny
classical model, but simply that the phenomena of measurement
uncertainty and measurement disturbance in quantum physics also arise
within a classical model with simple additional assumptions (a bound
on the sharpness of classical measurements) that do not overthrow our
classical conception of the world. Therefore, according to the
previously stated notion of classicality, measurement uncertainty,
complementarity, noncommutativity, and measurement disturbance are not
viewed as intrinsically quantum mechanical phenomena, and so we must
search deeper.

Because the 3 sharpest measurements, on the 6 extremal macrostates
turn out to have identical statistics to elementary quantum states and
measurements, the toy-theory also mimics interference -- in spite of
it being a classical statistical theory. At the simplest level,
interference is our ability to reversibly evolve a quantum state such
as $|\Psi\> = (|0\> + |1\> )/\sqrt{2}$ to another state $|\Psi'\> =
|0\>$, where $\{|0\>, |1\>\}$ is an orthonormal basis for a
2-dimensional, qubit system (see appendix~\ref{Qubits} for
details). The evolution acts linearly on $|\Psi\>$ and is defined by
the transformations $|0\> \mapsto (|0\>+ |1\>)/\sqrt{2}$ and $|1\>
\mapsto (|0\>- |1\>)/\sqrt{2}$. Under such an evolution the
$|0\>$-component of $|\Psi\>$ is enhanced (constructive interference),
while the $|1\>$-component of $|\Psi\>$ is eliminated (destructive
interference) \cite{Dirac, Ballentine}. If we were only keeping track
of the measurement statistics of measuring in the basis $\{ |0\>,
|1\>\}$, then the above interference would be described by the
reversible evolution of measurement statistics $(\frac{1}{2},
\frac{1}{2})$ to/from $(1,0)$.

Such behaviour is perfectly mimicked within the classical toy
model. For example, the minimal uncertainty macrostate $\a$ gives rise
to $(1/2, 1/2)$ outcome statistics for the $B$ measurement, but it can
be converted in a reversible way (deterministically shuffle the cells
around) into any of the other minimal uncertainty macrostates. The
reversible transformation of the cells $(0,1) \leftrightarrow (1,0)$ transforms
$\a$ into $\b$ and hence, the measurement statistics of $B$ go from
$(1/2,1/2)$ to $(1,0)$. If we simply look at the measurement
statistics, this is indistinguishable from the quantum example of
interference described above. Again, it is important to note that we
are \emph{not} claiming that general quantum interference is
classical, merely that a similar phenomenon can appear in the
reversible dynamics of classical statistical models and so a simple
notion of ``interference'' is perhaps too blunt an answer to the
question of what quantum phenomenon is intrinsically non-classical,
and so requires more precision.

Remarkably, the above classical toy-model exhibits an array of other
phenomena traditionally associated with quantum mechanics, such as the
collapse of the wave-packet, state teleportation \cite{teleportation}
and the impossibility of cloning states \cite{noclone1, noclone2}. All
the striking phenomena stem solely from a single restriction on
classical resolving power. Instead of describing all these within the
toy-model, we shall upgrade the basic RR-condition to obtain a more
natural, and intuitive scenario where these phenomena are more vivid
and generate a genuine classical fragment of quantum theory.

\subsection{Gaussian states and operations are a classical fragment of
  quantum physics}

\label{Gaussian}
 
The toy model of a classical particle in a box, subject to a single
constraint on resolving power, leads to a mimicking of phenomena that are
traditionally deemed quantum mechanical in character. However, this simple
classical model can now be magnified into a more surprising result
that has direct contact with quantum mechanics \cite{ERL-mech}.
 
Consider the phase space of a classical system \cite{Goldstein},
parameterized by variables $(\x, \p)$, but now imagine discretizing
the phase space into boxes with sides of length $L$ and imposing a
Resolution Restriction condition such that our Liouville distributions
can never have smaller support than some limited number of boxes over
the phase space. Such a simplistic approach would be clunky in that it
would depend on an arbitrary way of partitioning phase space. To get
around this we can instead impose an RR-constraint at the level of the
expectation values of canonical coordinates. When we do so, we find
that the resulting physical theory makes exactly the same predictions
as \emph{Gaussian quantum mechanics}\footnote{We give a short account
  of Gaussian states and operations in appendix~\ref{CVS}. See
  \cite{GaussianReview} for a more in depth review.}.

For simplicity, we restrict attention to a single classical particle
moving in one spatial dimension, but the construction can easily be
generalized to any classical system. The particle has microstates
$(x,p)\in \mathbb{R}^2$ that make up the system's phase space. A
statistical Liouville distribution is then a probability distribution
$f$ on the phase space with $f(x, p) \ge 0$ and such that $\int \!\!
\d x \d p \, f( x, p) =1$.  From this we can compute the statistical
properties of the system, such as the expectation value of position
$\<x\> = \int \!\! \d x \d p \, [f(x,p) x]$, or the expectation value
of momentum $\<p\>:= \int \!\!  \d x \d p \, [f(x,p) p]$.

In classical mechanics there is no limit on how sharp the predictions
of $f$ can be -- we can know the precise microstate to any degree of
accuracy. An RR-constraint can be imposed by limiting how small the
\emph{fluctuations} about the mean $(\<x\>, \<p\>)$ can be.  The form
that this constraint should take is motivated by the symmetries and
structure of phase space, e.g.\ we want the constraint to be preserved
by classical time evolution of the system.  Since classical dynamics
may cause fluctuations in $x$ to be transferred into fluctuations in
$p$ and vice versa, and may induce correlations between $x$ and $p$,
it is natural to impose the constraint on fluctuations in $x$, $p$,
and their correlations taken all together. To do this we construct a
\emph{fluctuation matrix} $\gamma$, which is given in terms of
position and momentum by
\begin{equation}
  \gamma = \begin{bmatrix} 
    (\Delta x)^2& \<x p\> - \<x\>\<p\> \\
    \<x p \> - \<x\> \<p\>& (\Delta p)^2
  \end{bmatrix},
\end{equation}
where $(\Delta x)^2 = \<x^2\> - \<x\>^2$ and $(\Delta p)^2 = \<p^2\> -
\<p\>^2$ are the variances of the position and momentum in the
particular Liouville distribution $f$.

Using the matrix $\gamma$, we can define an RR-condition that
restricts how small the fluctuations in $f$ can get. An elegant way to
do this is to demand that $\gamma$ obey the matrix equation
\begin{equation}
  \label{contrr}
  \gamma + \lambda C \ge 0,
\end{equation}
for some $2\times 2$ matrix $C$, and some constant scale $\lambda \geq
0$ that measures the size of the ``boxes'' on phase space, where
Eq.~\eqref{contrr} means that the eigenvalues of $\gamma + \lambda C$
are all $\geq 0$.  Since the eigenvalues of $\gamma$ itself are always
$\geq 0$, the case $\lambda=0$ corresponds to switching off the
RR-constraint, and so $\lambda$ controls the level of fluctuations
within the toy classical universe.

The question of which matrix $C$ to use is more subtle, but we want to
choose it such that Eq.~\eqref{contrr} is preserved under classical
time evolution, i.e.\ if it holds at time $t$ then it should also hold
at ant time $t' > t$ under any dynamics allowed by classical
mechanics.  It turns out that setting $C = i \Sigma$, where $i$ is the
imaginary unit and
\begin{equation}
  \Sigma = \begin{bmatrix}
    0 & -1 \\
    1 & 0 \\
  \end{bmatrix},
\end{equation}
does the trick, so the final RR-condition is
\begin{equation}
  \label{RR-gaussian}
  \gamma + i \lambda \Sigma \ge 0,
\end{equation}
for some fixed minimal resolving scale $\lambda$ on the classical
phase space.  See appendix~\ref{RR} for more details of this
construction.

Finally, because we are following a statistical mechanics account of
the physics \cite{Tolman, Jaynes}, for a given fluctuation matrix
$\gamma$, we use the Gibbsian distribution $f$ that maximizes the
thermodynamic entropy $S = -\int\!\! \d x \d p \, f(x,p) \log
f(x,p)$. Thus the scenario we have described is precisely one of
classical statistical mechanics where our classical resolving power is
bounded in phase space by a scale $\lambda$. Indeed,
Eq.~\eqref{RR-gaussian} implies that $\Delta x \Delta p \ge \lambda$,
and so we find that the RR-condition encodes a classical uncertainty
relation on the statistical system.

The RR-condition can be interpreted as a kind of externally imposed
complementarity between position and momentum for the classical
system, and, as with the previous toy model, we have
measurement-disturbance: localizing the position of the particle must
randomly disturb its momentum in order to maintain repeatability and
the RR-condition. Once again, the theory displays a type of
``interference'' in terms of its macrostates in the sense described
earlier, and the structure of the extremal macrostates of the theory
is surprisingly rich.

To what degree does this mimic quantum mechanical phenomena?  What
classical fragment of quantum theory does this model define? The
answer is perhaps surprising -- the above scenario, in which we use
classical statistical mechanics under the RR-condition
Eq.~\eqref{RR-gaussian}, is precisely isomorphic to Gaussian quantum
mechanics \cite{ERL-mech, GaussianReview} (see appendix~\ref{CVS} for
a definition of Gaussian quantum mechanics). In other words,
\begin{center}
  \fbox{\parbox{0.95\textwidth}{\emph{If Nature only prepared Gaussian
        quantum states, and only performed Gaussian evolution and
        Gaussian measurements, then classical statistical mechanics
        with a single resolving constraint (with $\lambda =
        \frac{1}{2}\hbar$) would \emph{perfectly} reproduce all
        physical predictions. }}}
\end{center}
The proof that all of the features of Gaussian quantum mechanics are
reproduced is an involved computation, and we refer the reader to
\cite{ERL-mech} for the details.

Thankfully, there more to life than Gaussian physics, but this
correspondence still tells us some useful things. Firstly it shows
that Liouville mechanics under the RR-condition reproduces all quantum
phenomena that are present in Gaussian quantum mechanics
\cite{GaussianReview}.  This includes teleportation, superdense coding
\cite{densecode}, remote steering \cite{steering}, secure-key
distribution \cite{BB84}, no-cloning \cite{noclone1, noclone2}, and
the collapse of the wave-packet (we discuss some of these
shortly). However, the real value of the result is that it narrows our
hunt and tells us that the intrinsically non-classical phenomena, such
as Bell non-locality \cite{Bell-EPR, CHSH}, quantum computation
\cite{Nielsen-Chuang} and contextuality \cite{Kochen-Specker}, must
necessarily be \emph{non}-Gaussian in nature.

It should also be emphasized that in a precise sense there is no
``middle-ground'' between Gaussian quantum mechanics and the full set
of quantum operations. Specifically, the set of unitary
transformations that describe the dynamics in Gaussian quantum
mechanics are all those of the form $e^{-iHt}$, where the Hamiltonian
$H$ is at most quadratic in the canonical coordinates $q$ and $p$.
Now imagine that we have a single non-quadratic term $H_\star$ that
can be added to the Hamiltonain and which can be switched on or off
whenever we want. By adding this single $H_\star$ to the set of
quadratic Hamiltonians, the set of unitary transformations we can
achieve explodes to become the \emph{full set} of unitaries on the
Hilbert space \cite{ContinuousReview}. This means that the Gaussian
fragment is in a sense the largest classical fragment we can obtain by
following this line, and must rest right up against genuine
non-classicality.

\subsection{No-cloning is a classical statistical phenomenon}

\label{Cloning}

It is often maintained that the impossibility to clone quantum
information is a distinctly quantum mechanical phenomenon. Formally,
the no-cloning theorem \cite{noclone1, noclone2} says that it is
impossible to construct a physical device that, on input of any
quantum state $|\Psi\>$ will return the duplicated state $|\Psi\>
\otimes |\Psi\>$. Indeed, if such a magical device existed then one
could even violate relativity and signal faster than
light\footnote{The rough idea is that if Alice repeatedly clones one
  half of an entangled state that has been collapsed by a remote
  measurement made by Bob on the other half of the state, then she can
  magnify the information as to what type of state she possesses
  (e.g.\ whether it is a momentum eigenstate or a position
  eigenstate). Bob can use this to signal faster than light by
  choosing which type of state to collapse Alice's system to (e.g.\
  momentum eigenstates=``yes'', position eigenstates=``no'').}.

The proof of the no-cloning theorem in quantum theory is very
straightforward. Suppose a device existed that could clone two
nonorthogonal and nonidentical states $|\Psi\>, |\Phi\> \in \H_s$,
where $\H_s$ is the Hilbert space of the system.  Any physically
allowed transformation in quantum theory is described by a unitary
operation $U$ on the joint Hilbert space $\H_s \otimes \H_a$ of the
primary system and some apparatus system, which together form a closed
system. For a device that clones $|\Psi\>$ and $|\Phi\>$, this
transformation must satisfy
\begin{align}
  \label{clonedef}
  U(|\Psi\> \otimes |\Xi\>) & = |\Psi\> \otimes |\Psi\> \\  U(|\Phi\>
  \otimes |\Xi\>) & = |\Phi\> \otimes |\Phi\>,
\end{align} 
where $|\Xi\>$ is the initial state of the apparatus.  We can now
compute the inner product of the output states $U|\Psi\> \otimes
|\Xi\>$ and $U|\Phi\> \otimes |\Xi\>$ in two different ways.
Firstly, using Eq.~\eqref{clonedef}, we have
\begin{equation}
 | \< \Phi | \otimes \< \Xi| U^\dagger U | \Psi\> \otimes |\Xi\>| =
 | \< \Phi | \Psi \> \< \Phi| \Psi \>| = |\< \Phi | \Psi \>|^2.
\end{equation}
Alternatively, we can use the fact that $U$ is a reversible unitary
evolution and so $U^\dagger U = \I$, where $\I$ is the identity
operator, to give
\begin{equation}
 | \< \Phi | \otimes \< \Xi| U^\dagger U | \Psi\> \otimes |\Xi\>| =
  |\< \Phi | \Psi \> \< \Xi| \Xi \> |= |\< \Phi | \Psi \>|,
\end{equation}
where the last line follows from the fact that $| \Xi \>$ is a unit
vector.  In other words, a unitary transformation preserves inner
products.  Equating the two expression gives $|\<\Phi|\Psi\>| =
|\<\Phi|\Psi\>|^2$, which is only satisfied when $|\<\Phi| \Psi\>|$ is
either $0$ or $1$.  However, since we assumed that $|\Psi\>$ and
$|\Phi\>$ are nonidentical and nonorthogonal, this is a contradiction,
and thus there is no physically allowed cloning device in quantum theory.

Whilst classical bits can be freely copied, this only applies to the
exact values of the bits themselves.  In light of the theories
presented so far, it is perhaps better to think of a pure quantum
state as analogous to a probability distribution over the values of
classical bits, and these cannot be cloned.  We find that the
imposition of the RR-condition on the above classical statistical
model generates precisely the same prohibition as in quantum theory:
it is impossible to build a cloning device within this classical
theory.

The proof parallels the quantum proof, and relies only on a basic
property of Hamiltonian dynamics \cite{ERL-mech}. Suppose a classical
device existed that could clone two overlapping but nonidentical
Liouville distributions $f_S(\x,\p)$ and $g_S(\x,\p)$ defined on a
classical system $S$.  Let $C$ be a second classical ``apparatus'' of
equal size to $S$ that is initialized in some fiducial macrostate
$h_C(\x',\p')$. The composite system $SC$ undergoes some Hamiltonian
dynamics on the underlying microstates, which is assumed to clone
$f_S$ and $g_S$. The initial joint state is thus either $f_{SC} ( \x,
\p ; \x' , \p') = f_S (\x, \p) h_C (\x' , \p')$ or $g_{SC} ( \x, \p ;
\x' , \p') = g_S (\x, \p) h_C (\x' , \p')$, where $f_S $ and $g_S$ are
the input macrostates of the system, and $f_{SC}$ and $g_{SC}$ are
defined on the joint phase space $\mathbb{R}^{2N} \times
\mathbb{R}^{2N}$. If this evolution is a cloning process then, under
the Hamiltonian dynamics, we must have
\begin{align}
  \label{classclonedef}
  f_{S}( \x, \p)h_C ( \x', \p') & \longrightarrow f_S(\x, \p) f_C
  (\x', \p') \\
  g_S(\x,\p) h_C(\x,\p') & \longrightarrow g_S(\x, \p) g_C (\x', \p'),
\end{align}
where $f_S$ and $f_C$ are identical distributions, and similarly for
$g_S$ and $g_C$.  
 
To mirror the inner product computation of quantum states, we use a
classical measure \cite{GeometryQuantumStates} of how much two
statistical distributions overlap. The only fact we need about the
dynamics is the following: if $f$ and $g$ are two distributions on
phase space, then the overlap integral $F(f, g)=\int \d \x \d \p
\sqrt{f (\x, \p)} \sqrt{g (\x, \p)} $ is constant in time. Here, $F$
is the classical ``fidelity'' measure, quantifying the degree to which
the distributions overlap -- if $f=g$ then they overlap fully and
$F=1$, while if they have zero overlap then $F=0$. Put another way,
Hamiltonian dynamics evolves phase space distributions in a
volume-preserving way, similar to an incompressible fluid, which
implies that the fidelity is also preserved.
 
Now, we can compute the overlap of the two final states in two
different ways.  Firstly, using Eq.~\eqref{classclonedef}, we have
\begin{equation}
  F(f_Sf_C, g_S g_C) = F(f_S, g_S) F(f_C, g_C) = F(f_S, g_S)^2.
\end{equation}
However, since Hamiltonian dynamics preserves fidelity, we can
alternatively compute the fidelity of the initial states, which is
\begin{equation}
  F(f_Sh_C, g_Sh_C) = F(f_S, g_S) F(h_C, h_C) = F(f_S, g_S).
\end{equation}
In parallel to the quantum case, these two equations can only be
satisfied if $F(f_S, g_S) = 0$ or $1$, but, since we assumed $f_S$ and
$g_S$ are overlapping and non-identical, this is a contradiction.  We
conclude that it is fundamentally impossible to construct a device
that clones overlapping statistical distributions within the classical
theory.  Therefore the no-cloning theorem does not only apply to
quantum theory, but also to classical statistical mechanics.  Hence,
it should not be considered an intrinsically quantum mechanical
phenomenon.

\subsection{The EPR argument and the collapse of the wave-packet}

\label{EPRsec}

The seminal 1935 paper \cite{EPR} by Einstein, Podolsky and Rosen
asked whether quantum theory is ``complete'' or ``incomplete''. In
other words, perhaps quantum theory is only a stop-gap and there is a
yet deeper theory, in which ``God does not play dice''. Their analysis
revolved around a two particle state, which displays correlations
between both the position and momentum degrees of freedom.

It turns out that the EPR state and the measurements they considered
actually lie within the classical fragment we have just described --
in other words, at least as far as the EPR argument and measurements
are concerned, the answer is a tentative ``yes'': a perfectly
deterministic and local completion does exist that reproduces all the
statistics of the EPR experiment.  It is only when we get to Bell's
theorem that entanglement prohibits any such underlying degrees of
freedom that only interact locally, and thus the classical picture of
the world becomes untenable.
 
The EPR two-particle state is correlated in the two position degrees
of freedom and has zero total momentum. Specifically, in the position
representation, the state is
\begin{equation}
  \label{EPR}
  |\Psi_{1,2}\> = \int \d x_1 \d x_2 \, \delta(x_1 - x_2 + c) | x_1\>_1 |x_2\>_2,
\end{equation}
where $\delta(x)$ is the Dirac delta-function distribution and $c$ is
a constant.  In the momentum representation the same state is
\begin{equation}
  \label{EPRp}
  |\Psi_{1,2}\> = \int \d p_1 \d p_2 \, \delta(p_1 + p_2) | p_1\>_1 |p_2\>_2.
\end{equation}
Such a state might be obtained through the decay of some massive
particle with zero momentum into two lighter particles that propagate
in opposite directions and are now a distance $c$ apart\footnote{Note
  that, although the delta functions make this state unphysical, the
  same argument can be run with properly normalized Gaussian states
  that approximate them \cite{ERL-mech}.  We use the idealized version for
  simplicity.}.

Now, according to the orthodox account of quantum theory (i.e.\ the
textbook treatment of quantum mechanics), the state $|\Psi_{1,2}\>$
represents a situation in which neither particle has a definite
position, since it is not an eigenstate of either of the operators
$\hat{x}_1$ or $\hat{x}_2$.  Similarly, neither particle has a
definite momentum.  All we can say is that the system is in a state of
definite \emph{relative} position and momentum, i.e.\ there is perfect
correlation between the two positions and momenta.  Note that, in the
orthodox account, it is not simply a matter of each particle having a
definite position and momentum that is currently unknown to us, but
rather that the individual positions and momenta do not exist, since
the only properties that can be ascribed to the system are those
corresponding to operators of which the state is an eigenstate.

The EPR argument amounts to the observation that if one measures the
position of the first particle to be $x$ then the state of the
remotely separated second particle is collapsed to a sharp position
state $|x+c\>_2$, which, if measured, will always yield the value
$x+c$ with certainty.  Thus, according to the orthodox account, the
position of the second particle pops into existence as soon as the
first is measured and, since $c$ is arbitrary, they may be arbitrarily
far apart.  This represents a kind of nonlocality in the orthodox
account, since an observation made over here can cause something to
instantaneously pop into existence very far away.  The only way to
avoid this is to ``complete'' quantum mechanics by positing that, in
fact, the second particle did have a definite position before the
first was measured, and this is exactly what EPR argued for.

The same argument can also be run in momentum space.  If one chooses
to measure the momentum of the first particle and finds it to be $p$,
then the state of the second particle is collapsed to a sharp momentum
state $|-p\>_2$.  Thus, according to the orthodox account, the
momentum of the second particle pops into existence upon measuring the
momentum of the first, and so EPR argued that the second particle must
have a definite value of momentum prior to measurement in order to
avoid nonlocality.  It is rather striking that, on the orthodox
account, the choice of which observable to measure affects which
property pops into existence at a distant location, and that the
completion of quantum mechanics proposed by EPR would violate a strict
interpretation of the uncertainty principle.  Note that this extra
wrinkle on the argument is not required to establish the nonlocality
of the orthodox account, which already follows just from considering
position measurements on their own.

It turns out that the statistical model \cite{ERL-mech} with
restricted resolving power, can exactly reproduce the EPR measurement
statistics for position and momentum since they are Gaussian
measurements. Within this parallel model the ``paradoxical'' or
potential conflict with relativity is illusory and no longer a
concern. To represent the state given in Eqs.~(\ref{EPR}, \ref{EPRp})
in the classical scenario, we define the two-particle distribution on
the phase space $\mathbb{R}^{4}$ given by
\begin{equation}
  f(x_1, p_1 ; x_2, p_2) = \delta(x_1 - x_2 + c) \delta(p_1 + p_2), 
\end{equation}
which is the limit of a sequence of Gaussian distributions that all
satisfy the RR-condition.  This distribution has perfect correlations
between the spatial and momentum degrees of freedom of the
particles. The account of the EPR experiment now takes on a simple
form: the actual state of the particles are microstates of definite
position definite momentum. When we perform a local measurement that
determines the position of the first particle it is simply the
\emph{probability} distribution for the total system that changes, and
not the physical state of the remote system. This is no different to
the example of the correlated coins that was provided in the
introduction, and highlights that in many ways the collapse of the
wave-packet is no more strange than the updating of a probability
distribution; the objective state of the remote system remains the
same, and there is no conflict with relativity theory. It requires
a stronger result such as Bell's theorem to fully and conclusively rule out an
account along these lines.

\subsection{Is $\Psi$ analogous to a thermodynamic Gibbs state?} 

\label{Gibbs}

Some people might claim that quantum mechanics is not the final
theoretical framework for physics -- that there might be some even
more fundamental theory yet to be discovered, which includes quantum
theory as some kind of limiting case. Could it really be that quantum
theory is incomplete \cite{EPR}, and that there are underlying
variables that give a more fine-grained description, and that our
measurement devices actually respond to these underlying variables?
Historically \cite{Tolman, Callen}, this is what happened with
thermodynamics and statistical mechanics -- the macroscopic properties
of heat, temperature and pressure are well-defined properties obeying
the laws of thermodynamics, however they admit a statistical
mechanical description in terms of the rapid motion of underlying
variables -- atoms. How do we know that something like this could not
happen in the future with quantum mechanics? Could the wave-function
be more like the Gibbs distribution of statistical mechanics, and
somehow point to new underlying degrees of freedom?

We shall see that this can never happen. In a precise sense, and under
a broad range of entirely reasonable assumptions, no such thing can
happen in \emph{any} future theory of physics. To tackle such a seemly
nebulous question, we must use a sufficiently general framework that
contains only the most primitive notions of ``states'' and
``statistical measurements'', and which can account for the
predictions of quantum theory as a special case. Since the framework
we describe can account for the predictions of quantum physics, as
well as theories which are \emph{not} quantum theory, the framework
effectively allows us to regard quantum mechanics as an object in
itself, and to delineate its properties in contrast to other theories,
including classical mechanics.

According to the textbook account \cite{Dirac, vonNeumann}, the
quantum state $|\Psi\>$ provides a full description of a quantum
system, both in terms of its subsequent evolution in time and how it
responds to any measurement that we may wish to perform.  Every
orthonormal basis of the Hilbert space $\H$ corresponds to a quantum
measurement, and has outcome probabilities given by the Born-rule.  A
\textit{qubit} is any quantum system whose Hilbert space is
two-dimensional and so any state of the system is expressible as
$|\Psi\> = \alpha | 0\> + \beta |1\>$ for an orthonormal basis
$\{|0\>,|1\>\}$ and complex numbers $\alpha, \beta \in \mathbb{C}$
obeying $|\alpha|^2 + |\beta|^2 = 1$. A general orthonormal basis
contains two states $\{ |\Phi\>, |\overline{\Phi}\>\}$, and when a
measurement is performed in this basis on a system prepared in the
state $|\Psi\>$, then the outcome probabilities are simply given by
$|\< \Phi|\Psi \>|^2 $ and $ |\< \overline{\Phi} |\Psi\>|^2$.

Given this framework, what would it mean for the quantum state
$|\Psi\>$ to be like a Gibbs state and admit some hypothetical
microscopic structure?  This would require that there exists some set
$\Lambda$ of perhaps more fundamental states that give a sharper
description of the system \cite{Harrigan1, Harrigan2}. If this did
turn out to be the case, then instead of using $|\Psi\>$ to describe
the physics we could replace it with a \emph{probability distribution}
over this set of variables $\Lambda$. More precisely, each quantum
state $|\Psi\>$ would be associated with a probability
distribution\footnote{Even more precisely, probability distributions
  should be associated to the procedures for preparing quantum states
  rather than the states themselves to account for a subtlety called
  preparation contextuality \cite{Spekkens-contextuality}.  However,
  this subtlety does not affect any of the results presented here.
  See \cite{MattReview} for a more rigorous treatment that does deal
  with this topic.}  $p(\cdot|\Psi) : \Lambda \rightarrow \mathbb{R}$
for which $p (\lambda |\Psi) \ge 0$ for any point $\lambda \in
\Lambda$ and which is normalized as
\begin{equation}
  \int_\Lambda \!\! \d \lambda \, p (\lambda |\Psi) = 1.
\end{equation}
In particular, this means that the integral
\begin{equation}
  \int_R \d \lambda \, p (\lambda |\Psi )  = P(R|\Psi)
\end{equation}
is the probability that the underlying microstate state of the system
is in the region $R$ of the full state space $\Lambda$ when the state
$|\Psi\>$ is prepared experimentally.

Once we have a notion of quantum states being described by probability
distributions, how do we describe quantum measurements?  This too is
relatively easy. Suppose we perform a measurement in the basis $\{
|0\>, |1\>\}$ on a qubit prepared in the quantum state $|\Psi\> =\cos
\theta |0\> + \sin \theta |1\>$. According to quantum theory, the
probabilities of the two outcomes are $q_0 = \cos^2 \theta$ and $q_1
=\sin^2 \theta$. However, in terms of the underlying variables, the
measurement is described by a conditional probability distribution
$p(j|\lambda)$, where $p(j| \lambda)$ is the probability that the
measurement will return the $j^{\rm th}$ outcome when the system
occupies the microstate $\lambda$.  This means that we must have
$p(j|\lambda) \ge 0$ for all $\lambda \in \Lambda$, and also
$p(0|\lambda) + p(1|\lambda) =1$, so that the probabilities sum to
one. If these functions are to correctly describe the observed
measurement statistics then they must obey:
\begin{equation}
  \label{reproduce}
  \int_\Lambda\!\! \d \lambda \, p(j|\lambda) p(\lambda| \Psi) = q_j
\end{equation}
for $j=0,1$.

More generally, for systems of arbitrary dimension, a measurement in
the basis $\{ |\Phi_0\>, |\Phi_1\>, \dots |\Phi_{n-1}\>\}$ performed
on a system prepared in the state $|\Psi\>$, is described by a
conditional probability distribution $p(\Phi_j|\lambda)$ over the $n$
outcomes that obeys
\begin{equation}
  \label{reproducegen}
  \int_\Lambda\!\! \d \lambda \, p(\Phi_j|\lambda) p (\lambda |\Psi) =
  |\<\Phi_j |\Psi\>|^2 
\end{equation}
for all $j=0, \dots , n-1$. This is what would be required in any
hypothetical theory in order for the quantum predictions for
measurement outcomes to be reproduced.

It is vital to emphasize that this formulation \emph{includes} the
orthodox description of quantum mechanics, and so it is a broader
framework that allows questions to be posed that are impossible within
the traditional setting. To show that it includes the orthodox
account, simply take $\Lambda$ to be the set of all quantum states
$\Lambda =\mathcal{Q} =\{|\Psi\>\}$, where states that differ by a
global phase are identified, and let $p(\lambda|\Psi)$ be a delta
function distribution $p(\lambda |\Psi) = \delta (\lambda - \Psi)$,
with weight just on the quantum state $| \Psi \>$ that is
prepared. The measurement conditional probabilities are then simply be
taken to be $p(\Phi_j|\Psi) = |\<\Phi_j | \Psi\>|^2$, to trivially
recover the Born-rule.

Why then should we bother to use such a description? The reason is
that since this is a general, probabilistic setting that only requires
the notion of abstract states $\lambda \in \Lambda$ and probability
distributions, it uses only the most elementary notions of what one
normally calls a ``physical theory''. Such breadth makes it powerful,
and will allow us to rule out alternative theories, identify
intrinsically quantum phenomena, and to study quantum theory ``from
the outside''.

\subsection{Overlapping distributions: The
  Kochen-Specker model}

\label{KS}

In the orthodox account of quantum theory in which $p(\lambda |\Psi) =
\delta (\lambda - \Psi)$, the distributions corresponding to different
quantum states do not overlap -- they are simply delta functions
located at the different quantum states.  However, for a qubit, there
is a slightly more interesting representation, due to Kochen and
Specker, which can be used to frame a fundamental question concerning
the objectivity of the wave-function in quantum mechanics
\cite{Harrigan1, Kochen-Specker}.

An arbitrry quantum state of a qubit can be written as
\begin{equation}
  |\Psi\> = \cos \left ( \frac{\vartheta}{2} \right ) |0\> + e^{i
    \varphi} \sin \left ( \frac{\vartheta}{2} \right ) |1\>,
\end{equation}
where $0 \leq \vartheta \leq \pi$ and $-\pi \leq \varphi \leq \pi$.
By defining the vector $\boldsymbol{\Psi} = (\sin \vartheta \cos
\varphi, \sin \vartheta \sin \varphi, \cos \vartheta)$, we see that
the set of states of a qubit can be represented as points on the unit
sphere $S^2$, which is known in this context as the \emph{Bloch
  sphere} (see appendix~\ref{Qubits} for further details).  There is a
one-to-one relation between every qubit state $|\Psi\>$, and a vector
$\boldsymbol{\Psi}$ on the surface of the Bloch sphere and, in what
follows, we simply use this vector to represent the quantum state.
Moreover, if a state $|\Psi\>$ is measured in a basis $\{|\Phi\>,
|\overline{\Phi}\>\}$ then the probability of obtaining the $|\Phi\>$
outcome is
\begin{equation}
  \label{BlochBorn}
  |\<\Phi|\Psi\>|^2 = \frac{1}{2} \left ( 1 + \boldsymbol{\Phi} \cdot
    \boldsymbol{\Psi} \right ).
\end{equation}

\begin{figure}[htb]
  \begin{center}
    \includegraphics[width=8cm]{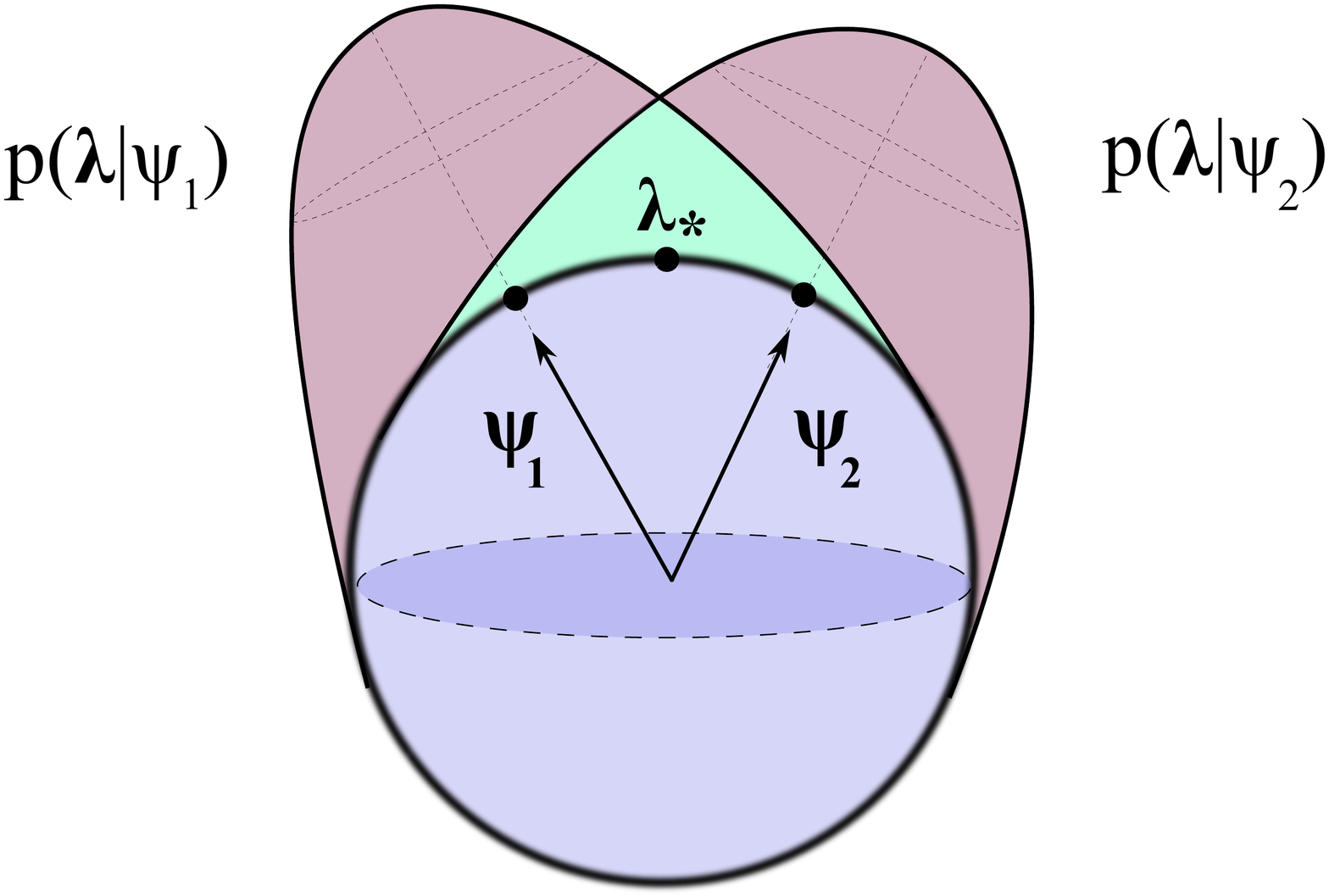}
  \end{center}
  \caption{\label{evil-rabbit}\textbf{The Kochen-Specker
      construction:} The set of states within the Kochen-Specker
    construction for a single qubit is simply the surface of a
    sphere. We plot $p(\boldsymbol{\lambda} |\Psi_1)$ and
    $p(\boldsymbol{\lambda} |\Psi_2)$ (the peaked, shaded caps with
    axial symmetry) for two non-orthogonal, pure quantum states
    $|\Psi_1\>$ and $|\Psi_2\>$, each of which is specified uniquely
    by their respective Bloch vectors $\boldsymbol{\Psi_1}$ and
    $\boldsymbol{\Psi_2}$. The probability distributions overlap, and
    so a microstate in the overlap-region (denoted
    $\boldsymbol{\lambda}_\star$ in figure) can equally be associated
    to either quantum state. If such a phenomenon were to occur for
    $\mbox{dim}(\H) >2$ within some future theory, then the
    wave-function $\Psi$ need not be an intrinsic property of an
    individual physical system. }
\end{figure}

Kochen and Specker's model for a qubit employs the unit sphere
$\Lambda = S^2$ as its space of microstates, and to every quantum state
$|\Psi\>$, they associate a probability distribution
\begin{align}
  p (\boldsymbol{\lambda} |\Psi) & = \frac{1}{\pi} \Theta
  (\boldsymbol{\lambda} \cdot \boldsymbol{\Psi}) \boldsymbol{\lambda}
  \cdot \boldsymbol{\Psi},
\end{align}
where $\Theta$ is the Heaviside step function
\begin{equation}
  \Theta(x) = \begin{cases} 1, & x \geq 0 \\
  0, & x < 0. \end{cases}
\end{equation}
In the Bloch sphere representation, this means that
$p(\boldsymbol{\lambda} |\Psi)$ is only nonzero if the angle $\theta$
between $\boldsymbol{\Psi}$ and $\boldsymbol{\lambda}$ is less than
$\pi/2$, and it takes the value $\cos \theta$ on this hemisphere.  See
Figure~\ref{evil-rabbit} for an illustration of these probability
distributions.

To represent measurements, a basis vector $|\Phi\>$ is represented by
the conditional probabilities $p (\Phi|\boldsymbol{\lambda}) = \Theta
(\boldsymbol{\lambda} \cdot \boldsymbol{\Phi})$, which describes the
probability of getting the outcome $|\Phi\>$ when the exact microstate
is $\boldsymbol{\lambda}$. On the Bloch sphere, this means that the
outcome will be $|\Phi\>$ if the angle between $\boldsymbol{\lambda}$
and $\boldsymbol{\Phi}$ is $\leq \pi/2$ and will be the orthogonal
basis state otherwise.  A direct calculation (e.g.\ see
\cite{MattReview}) shows that the model yields
\begin{equation}
  p(\Phi | \Psi) = \int \!\! \d \, \boldsymbol{\lambda} p(\Phi
  |\boldsymbol{\lambda}) p(\boldsymbol{\lambda} | \Psi) =
  |\<\Phi|\Psi\>|^2, 
\end{equation}
and so does in fact reproduce the Born rule for a qubit system.

Why is this construction of interest to us? Firstly, the
Kochen-Specker construction exactly reproduces the Born-rule for a
two-dimensional system and so can be viewed as mini classical fragment
of quantum theory.  However, the real reason it is of interest is that
if we plot the distribution functions for two different states we
notice a distinct feature of the representation -- two
\emph{different} quantum states $|\Psi_1\>$ and $|\Psi_2\>$ can have
\emph{overlapping} distributions. The core significance of this is
that there is a region of microstates (the light-shaded part of
Figure~\ref{evil-rabbit}) that belong to both $p(\boldsymbol{\lambda}
|\Psi_1)$ and $p(\boldsymbol{\lambda} |\Psi_2)$. If the system
occupies the microstate $\boldsymbol{\lambda}_\star$, which lies in
this overlap region, then a unique wave-function cannot be associated
to it. In other words, within such a hypothetical model a quantum
state $|\Psi \>$ would not be an objective property that is ``carved
into'' the physical system!

Why might we want the distributions corresponding to different quantum
states to overlap?  Recall from \S\ref{toy}, that the extremal states
of the toy model overlap, as do the distributions in the restricted
Liouville mechanics presented in \S\ref{Gaussian}.  This overlap
naturally explains why the extremal states cannot be perfectly
distinguished, and why distributions cannot be cloned.  Essentially,
if two preparation procedures sometimes lead to the exact same
microstate, then the action of any physical device cannot depend on
which of the two preparation procedures was used whenever a microstate
in the overlap region happens to be occupied.  Therefore, the
probability of successfully distinguishing or cloning macrostates is
limited by the probability assigned to the overlap region.  The
overlapping distributions in the Kochen-Specker model explain why
qubit states cannot be perfectly distinguished or cloned in the same
way within this model.

This raises the question of whether such overlapping distributions
could ever occur in some future theory for general quantum systems,
and, if they could, then how much would we have to contort our
perspective of the world? If we are necessarily forced to adopt
something ridiculous or bizarre then we would conclude that this is
not the case -- the quantum state must indeed be carved into physical
systems, and must be an objective label of a quantum system.  This
would show that the explanations of quantum phenomena in terms
of overlapping distributions, despite intuitive appeal, must in fact be wrong.

\section{Genuinely Non-Classical Aspects of Quantum Physics}

\label{NC}
  
The previous results show that some fragments of quantum theory can be
reproduced in ways that do not violently clash with our classical
intuitions, but it is clear that we were being led into more and more
contrived models in order to capture more and more of the phenomena of
quantum theory.  The take-home message so far is simply that many
commonly touted phenomena -- intrinsic randomness, complementarity,
measurement-disturbance, no-cloning, collapse of the wave-packet,
etc. -- do not in themselves dramatically challenge our classical
notions, as they already appear within theories such as the
statistical mechanics model with the RR-condition presented above. In
addition, we have shown that the physics of a two-dimensional quantum
system can be reproduced by a statistical model in which the
probability distributions associated to different quantum states can
overlap.  This raises the question of whether the quantum-mechanical
wave-function is necessarily an objective property of a quantum
system.

We now describe results that reveal fundamentally \emph{non-classical}
phenomena, in the sense that any classical account underlying the quantum
phenomena must be rather contorted. We begin with the seminal result of
John Bell \cite{Bell-EPR}, and then move onto two more recent results that
provide further insights into the strangeness of quantum theory.  The
first of these is Hardy's theorem \cite{Hardy-baggage}, which shows that the set
$\Lambda$ of microstates must be infinitely large, even for a finite
dimensional system, and hence that such systems must contain an
infinite amount of information.  The second is the
Pusey--Barrett--Rudolph theorem \cite{PBR}, which shows that, under
reasonable assumptions, the quantum state must be an objective
property of an individual quantum system.

\subsection{Bell's theorem: Quantum physics violates local causality}

\label{Bell}

The departure of quantum mechanics from classicality was put into a
very sharp and powerful form by John Bell \cite{Bell-EPR, Peres}, who
showed that some aspects of quantum entanglement can never fit into
a model in which systems possess objective properties prior to
measurement and that also obeys a principle of locality.  Since the
result only depends on certain empirically observed predictions of
quantum theory, rather than the structure of the theory itself, any
future theory beyond quantum theory will be subject to the same
argument, so there can be no going back to a conception of the world
that is both classical and local.

The version of Bell's theorem we present here is due to Clauser,
Horne, Shimony and Holt \cite{CHSH}, and is the one most commonly used
in experiments.  To understand it, we need to explain both the
mathematical components and the physical concepts.  To avoid confusion, it is
helpful to separate these two parts, so we begin with the mathematics.

The easiest way to understand the mathematics of Bell's theorem is in
terms of a cooperative game, in which Alice and Bob are playing as a
team against Charlie.  Suppose that Alice and Bob are captured and
held captive by Charlie. Alice and Bob are told that the next morning
they will be placed into two separate interrogation rooms with no
possibility of communicating with each other.  They will each be asked
one of two possible yes/no questions.  We call the question that Alice
gets asked $x$ and the question that Bob gets asked $y$.  For
definiteness, we can imagine that both Alice and Bob's questions are
labelled $0$ and $1$, so that $x$ and $y$ are binary variables that
take values $0$ or $1$.  Let $A_x$ be the answer that Alice gives to
question $x$ and let $B_y$ be the answer that Bob gives to question
$y$.  To be released they must get their stories straight in a very
particular way.  If they are both asked question $1$ then their
answers, $(A_1, B_1)$, must obey $A_1\ne B_1$. In all other cases,
i.e.\ if $(x,y) = (0,0), (0,1)$ or $(1,0)$, they must provide answers
for which $A_x = B_y$.

Alice and Bob are told that they can spend the night together to
discuss their strategy for answering the questions.  They also have
access to devices for generating classical randomness -- for
definiteness suppose they have a set of dice with different weightings
-- which they may use to determine their strategy and which they may
also bring into the interrogation room with them.  They are assured
that Charlie will not eavesdrop on their discussions and, in fact,
that he will choose which questions to ask completely randomly by
flipping two separate coins.  What is the best strategy for Alice and
Bob to adopt that gives them the highest chance of being released?

\begin{figure}[htb]
  \begin{center}
    \includegraphics[width=10cm]{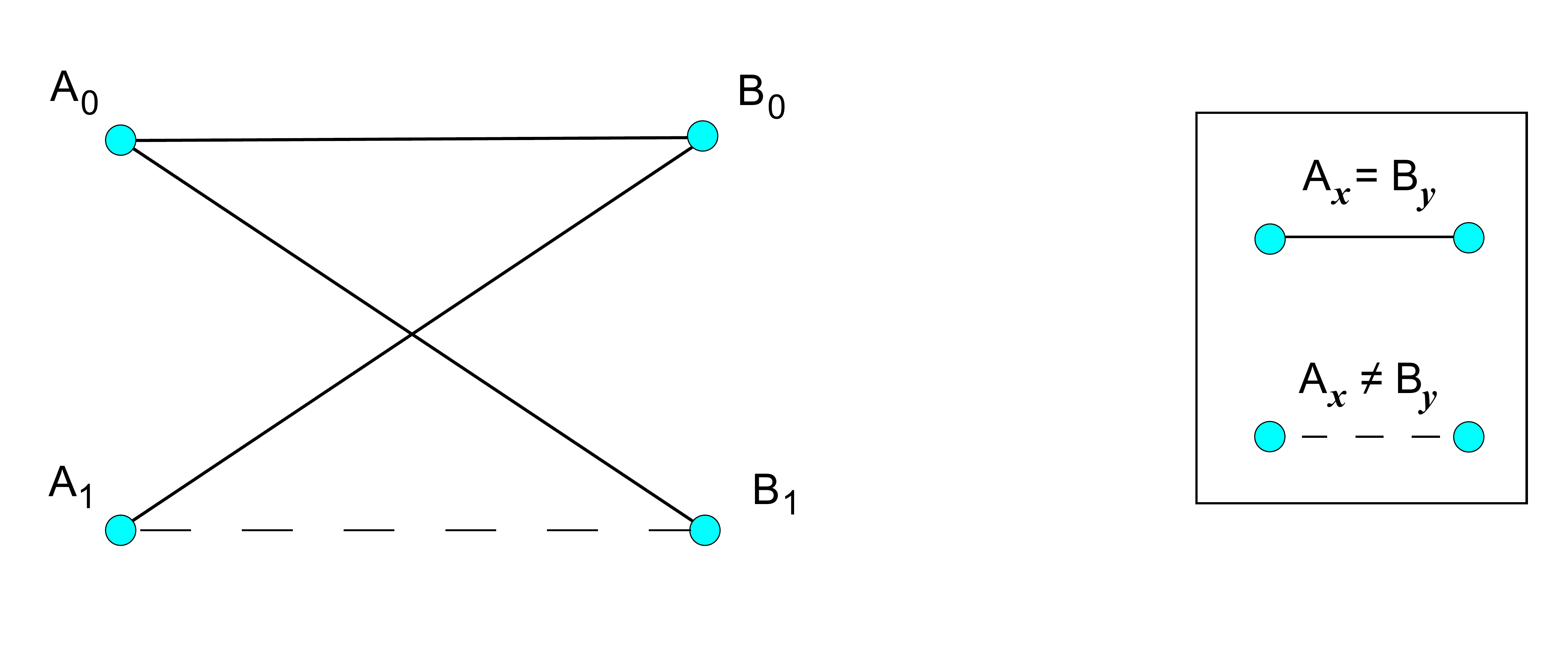}
  \end{center}
  \caption{\label{frustrated}\textbf{A Frustrated Network:} The
    winning answer patterns for Alice and Bob. A line denotes that
    Alice and Bob should try to give the same answer, while a dotted
    line denotes that Alice and Bob should try to give opposite
    answers. It is impossible to satisfy all $4$ constraints at the
    same time, but we can satisfy $3$ of them.}
\end{figure}

To begin with, let's ignore the possibility of using randomness and
ask what is the best that Alice and Bob can do if they employ a
deterministic strategy, i.e.\ they simply have to decide, in advance,
which answer Alice will give to her question and which answer Bob will
give to his.  It is helpful to represent the target answers as a graph
where the answers $A_0,A_1,B_0$ and $B_1$ are four vertices
\cite{Collins1, Collins2, Schmidt}, and we connect the variables that
should be equal by a solid line and those that should be different by
a dotted line (see Figure~\ref{frustrated}).

Now, let's traverse the links in the graph and try to satisfy as many
of the requirements as possible.  Suppose we start by setting $A_0$ to
``yes''.  Following the top solid line, we see that $B_0$ should equal
$A_0$, so it should also be assigned ``yes''.  The diagonal solid line
connecting $B_0$ to $A_1$ then implies that $A_1$ should also be
``yes''.  The dotted line from $A_1 $ to $B_1$ implies that $B_1$
should be different from $A_1$ so we set $B_1 = \text{``no''}$.
However, now we have a problem because the diagonal solid line from
$B_1$ to $A_0$ implies that these two should be equal, but we have
already set $A_0$ to ``yes'' and $B_1$ to ``no'', so we have only
managed to satisfy three of the four requirements.  Therefore, using
this strategy, Alice and Bob will win if Charlie picks any of the
question pairs $(0,0)$, $(1,0)$ or $(1,1)$, but they will lose if he
picks $(0,1)$.  Hence, their probability of winning is $3/4$, since
the chance of Charlie picking $(0,1)$ is $1/4$ if he chooses his
questions by two fair coin flips.

It is fairly easy to see that, however Alice and Bob assign ``yes''
and ``no'' to their questions, they can only satisfy at most three of
the four requirements.  This is because, however you go about
traversing the graph and assigning answers, you can never satisfy the
final requirement encoded in the final link because it contradicts the
implications of the other three.  Thus, $3/4$ is the largest
probability with which Alice and Bob can win the game via a
deterministic strategy.

This is the basic mathematics of Bell's theorem, but we still have to
deal with the possibility of probabilistic strategies that employ
randomness. There are two ways in which Alice and Bob can use
randomness.  The first is that, when they are still together the night
before the interrogation, they can roll some dice and each write down
the results.  They can then make their choice of answers depend on the
outcomes of the dice rolls.  For example, they might roll one dice and
agree that if it comes up odd then Alice and Bob should both answer
``no'' to all of the questions, whereas if it comes up even Bob will
switch his answer to question $0$ to ``yes''.  It will not make a
difference if Alice and Bob look at the dice roll outcomes while they
are still together and compute their answers, or if they simply write
down the outcomes of the dice rolls, take them with them to the
interrogation room, and perform the computation after the questions
are asked.  So long as they have agreed on a strategy for computing
the answers, this will have the same result.  It is easy to see that
this cannot increase their probability of winning the game.  On any
given outcome of the dice rolls, Alice and Bob will end up with some
specific set of answers, and the bound of $3/4$ will apply to these.
On average, they will win with probability $3/4$ whenever the dice
rolls lead to an assignment of answers that is an optimal
deterministic strategy, and with probability less than $3/4$ when they
do not.  Therefore, overall they may as well just pick an optimal
deterministic strategy to begin with.

The second way they can use randomness is to take some dice with them
into the interrogation room, roll them \emph{after} the questions have
been asked, and use the outcome to determine their answer based on a
pre-agreed strategy.  This looks like it adds generality because they
could choose to roll different dice, with different weightings,
depending on which questions they are asked.  For example, Alice might
take a green die and a blue die into the interrogation room with her
and roll the green one if she is asked question $0$ and the blue one
if she is asked question $1$.  In this case, Alice's answer does not
even really come into existence until the question is asked, so this
perhaps looks a bit more like what is going on in a quantum
measurement.

However, this cannot make a difference to the probability of winning
either.  Instead of waiting until she gets to the interrogation room,
Alice could just roll both the green die and the blue die while she is
still together with Bob.  Then, she could just write down both
outcomes and use one of them if she is asked question $0$ and the
other if she is asked question $1$.  This is already covered by the
first way of using randomness, where Alice and Bob can make their
answers an arbitrary function of the dice rolls they make when they
are together.  Although she is doing something physically different --
rolling two dice in advance vs.\ rolling just one die when she already
knows the question -- as far as the probability of winning the game is
concerned we might as well just move all the randomness generation to
the beginning, when Alice and Bob are still together, and we already
know this is no better than just choosing a deterministic strategy at
the outset.

Let us now reflect a bit on exactly what we have proved.  In general,
Alice and Bob can generate randomness when they are together and do
not know which questions they will be asked yet, and also separately
after they have been asked their questions.  Let's call the variables
that they generate when they are together $\lambda$, taking values in
some space $\Lambda$.  This will have some probability distribution
$p(\lambda)$.  If Alice and Bob were together and able to communicate
when they are asked their questions, then the most general thing they
could do would be to base their answers $A_x$ and $B_y$ on the
questions $x$ and $y$ they are asked, their prior shared randomness
$\lambda$ and any new randomness they choose to generate in the
interrogation room.  This can be described by conditional
probabilities $p(A_x,B_y|x,y,\lambda)$.

In general, a conditional probability distribution
$p(A_x,B_y|x,y,\lambda)$ can be decomposed as
\begin{equation}
  p(A_x,B_y|x,y,\lambda) = p(B_y|A_x,x,y,\lambda)p(A_x|x,y,\lambda). 
\end{equation}
What happens if we now take into account the fact that Alice and Bob
are separated and unable to communicate when they are asked their
questions?  Firstly, $A_x$ cannot depend on $y$, as Alice does not
know $y$ when she is asked her question, so we have
$p(A_x|x,y,\lambda) = p(A_x|x,\lambda)$.  Secondly, Bob does not know
$x$ or $A_x$ when he is asked his question, so $p(B_y|A_x,x,y,\lambda)
= p(B_y|y,\lambda)$.  Note that this does not mean that Bob has no
information about how Alice will answer her question.  In particular,
if they have chosen a deterministic strategy then Bob knows exactly
how Alice will answer each question.  However, the point is that
$\lambda$ already encodes all of the information Bob has about Alice's
strategy so, given $\lambda$, Bob's answer has no additional
dependence on $A_x$.

Altogether then, we have that
\begin{equation}
  p(A_x,B_y|x,y,\lambda) = p(A_x|x,\lambda)p(B_y|y,\lambda).
\end{equation}
This condition is known as \emph{local causality}, for reasons we
shall see shortly.

Finally, to work out the probabilities that Alice and Bob will answer
$(A_x, B_y)$ to the pair of questions $(x,y)$, we need to average over
the randomness $\lambda$ they generated to obtain
\begin{equation}
  p(A_x,B_y|x,y) = \int_{\Lambda} \d \lambda
  p(A_x|x,\lambda)p(B_y|y,\lambda)p(\lambda). 
\end{equation}

What we have proved via our long discussion is that, in any strategy
that satisfies local causality, Alice and Bob can win the game with
probability no greater than $3/4$, or, equivalently, if we define
$p(A_x = B_y) = p(\text{``yes''},\text{``yes''}|x,y) +
p(\text{``no''},\text{``no''}|x,y)$ and $p(A_x \neq B_y) =
p(\text{``yes''},\text{``no''}|x,y) +
p(\text{``no''},\text{``yes''}|x,y)$, we have
\begin{equation}
  \label{CHSH}
  \frac{1}{4} \left [ p(A_0=B_0) + p(A_0 = B_1) + p(A_1 = B_0) + p(A_1
    \neq B_1) \right ] \leq \frac{3}{4},
\end{equation}
which is an example of a Bell inequality.

Now, instead of classical randomness, let's see what happens if we
allow Alice and Bob to take correlated quantum systems with them into
the interrogation rooms, and to base their responses on the outcomes
of quantum measurements. Remarkably, this allows them to violate the
bound and win the game with probability $>3/4$.  Specifically, suppose
they share a pair of qubits in the singlet state
\begin{equation}
  |\Psi^-\>_{AB} = \frac{1}{\sqrt{2}} \left ( |0\>_A \otimes |1\>_B -
    |1\>_A \otimes |0\>_B \right ). 
\end{equation}
If Alice is asked question $0$ then she measures the Pauli observable
$\sigma_3$.  (If Alice's qubit is a spin-1/2 particle then the Pauli
observables $\sigma_1, \sigma_2$ and $\sigma_3$ correspond to the
angular momenta along the $x$, $y$ and $z$ directions
respectively. The Pauli observables are defined in
appendix~\ref{Qubits}.)  If the outcome is $+1$ she answers $A_0 =
\text{``no''}$ and if it is $-1$ she answers $A_0 =
\text{``yes''}$. If she is asked question $1$ then she instead
measures the Pauli observable $\sigma_1$ and answers
$A_1=\text{``no''}$ if she gets $+1$ and $A_1 = \text{``yes''}$ if she
gets $-1$.  Bob measures $(\sigma_3 + \sigma_1)/\sqrt{2}$ if asked
question $0$ and answers $B_0 = \text{``yes''}$ if he gets $+1$ and
$B_0 = \text{``no''}$ if he gets $-1$. If he is asked question $1$ he
instead measures $(\sigma_3 - \sigma_1)/\sqrt{2}$, and answers $B_1 =
\text{``yes''}$ if he gets $+1$ and $B_1 = \text{``no''}$ if he gets
$-1$.  A straightforward computation reveals that, with this strategy,
Alice and Bob will win with probability $(2 + \sqrt{2})/4 \approx
0.854$ \cite{Peres}.  This is strictly larger than the classical bound
and results from the super-strong correlations that exist within the
singlet state.

\begin{figure}[htb]
  \begin{center}
   \includegraphics[width=7cm]{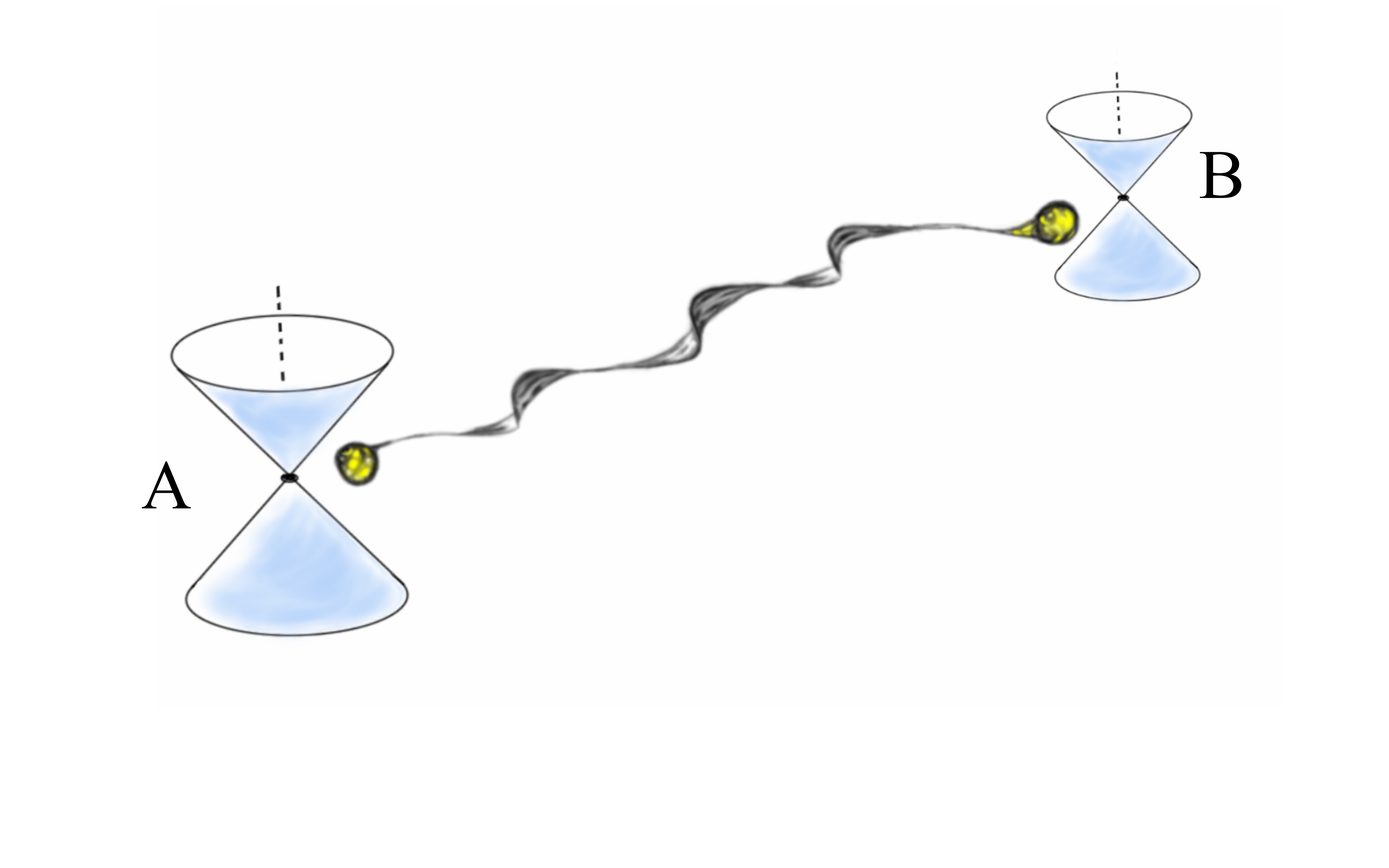}
    \includegraphics[width=7.5cm]{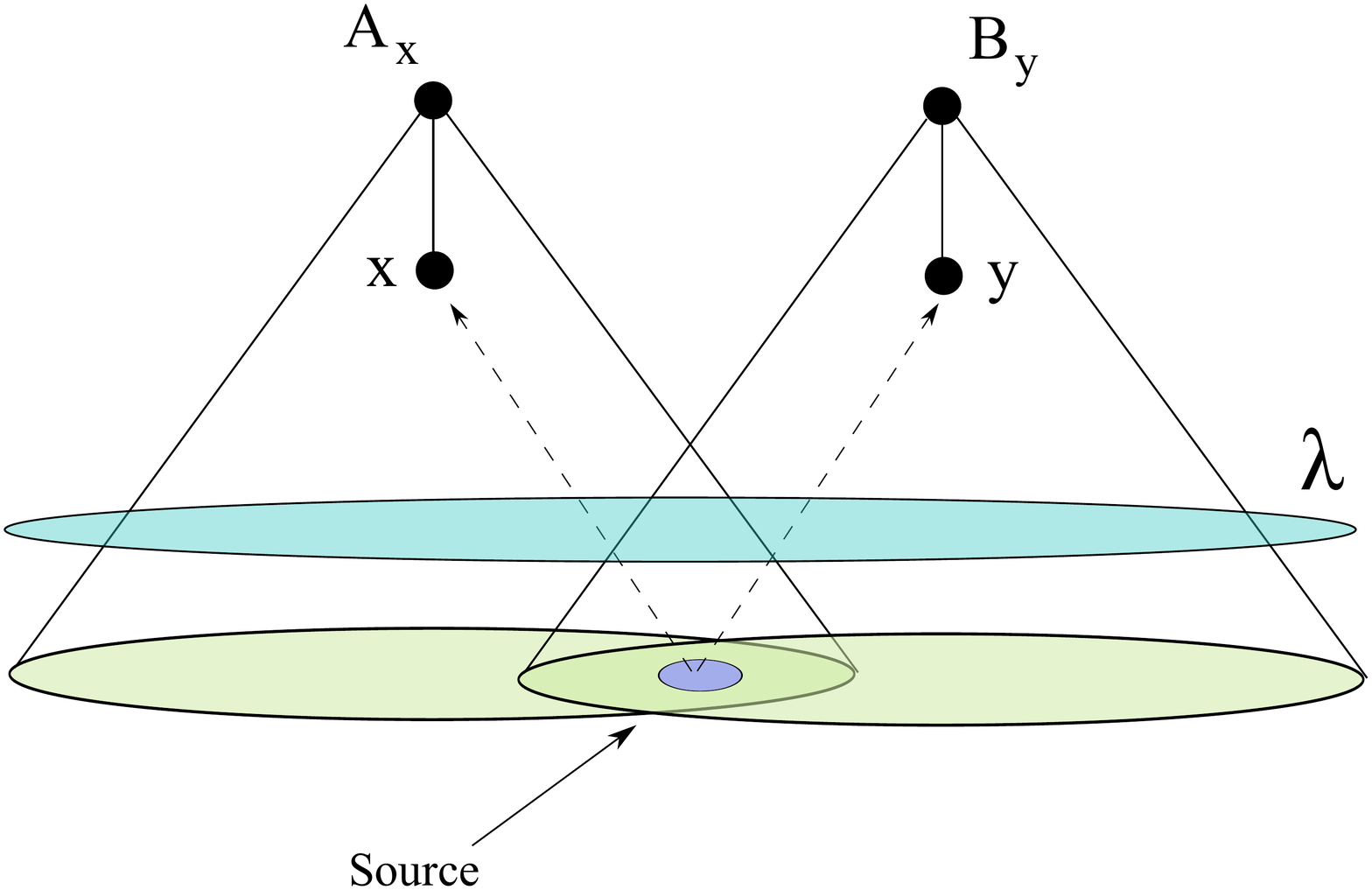}
  \end{center}
  \caption{\label{sketch}\textbf{The Violation of Local Causality:}
    Alice and Bob, who share an entangled quantum state (figure on the
    left), perform local quantum operations in their spacelike
    separated interrogation rooms. They are asked questions $(x,y)$
    and give answers $(A_x,B_y)$, with causal structure as shown
    (figure on the right). It is impossible for them to signal faster
    than light, but nonetheless local causality is violated.}
\end{figure}

This concludes our discussion of the mathematics of Bell's theorem,
but what does it all mean physically?  Consider the spacetime diagram in
Figure~\ref{sketch}.  A source emits two qubits in the singlet state,
which travel to two spacelike separated detectors, $A$ and $B$, where
an observable is measured on each of them.  Each of the detectors has
two settings, corresponding to the two questions that Alice and Bob
might be asked in the game.  We label the outcome at detector $A$ when
it has setting $x$ as $A_x$ and the outcome at detector $B$ when it
has setting $y$ as $B_y$.  The choice of which setting to use at each
detector is made at random, sufficiently late such that there is no
possibility of a signal from the point at which the choice of setting
$x$ is made, travelling at the speed of light or less, to the
detection event $B_y$ and similarly for $y$ and $A_x$.  If we set the
detectors to measure the observables described above, then we know we
can obtain the value $(2 + \sqrt{2})/4$ for the left hand side of
Eq.~\eqref{CHSH}, in violation of the inequality.

Given the causal structure of Figure~\ref{sketch}, what would we
expect to happen in a classical statistical model?  Consider a region
of spacetime that ``screens off'' $x$ and $A_x$ from $y$ and $B_y$ and
vice versa.  By this we mean that any timelike path from $x$ or $A_x$
to $B_y$ that passes through the source must travel through the
region.  One such region is highlighted in light blue in
Figure~\ref{sketch}, but many others are possible.  Let $\lambda$
denote the state of all of the fundamental degrees of freedom that
exist in this region. 

We would expect that any correlation that exists between the two wings
of the experiment should be mediated by $\lambda$.  This is because we
assumed that $\lambda$ describes \emph{all} of the fundamental degrees
of freedom in the region, so there is nothing else within the light
blue region that could possibly mediate correlations, and any causal
link between the two wings that does not pass through this region
would require super-luminal signalling.  In other words, we are in
precisely the same scenario that Alice and Bob face in their separate
interrogation rooms, unable to communicate with each other.  Thus, for
the exact same reasons as before, we expect local causality to hold,
i.e.\ $p(A_x,B_y|x,y,\lambda) = p(A_x|x,\lambda)p(B_y|y,\lambda)$.  As
we have already shown, this implies that the inequality given in
Eq.~\eqref{CHSH} should hold.

Remarkably, quantum violations of the inequality have been observed
experimentally in the experiments of Aspect et.\ al.\ \cite{Aspect},
and in numerous experiments since then \cite{Genovese, Pan,
  Brunner}. The implication of this is that local causality must fail,
and therefore either one accepts super-luminal influences at some
fundamental level, or that elementary notions of realism within
statistical theories must be discarded forever.  In the literature,
this is often referred to by saying that either ``locality'' or
``realism'' must be given up.  However you wish to parse the dilemma,
it is clear that Bell inequality violations imply a radical departure
from classical physics.

\subsection{Hardy's theorem: Quantum systems contain an infinite
  amount of information}

\label{Hardy}

At the heart of classical information theory is the idea of a
classical \textit{bit} -- the information revealed by a single yes-no
question. Our ability to quantify, encode and transform information
has revolutionised the world in countless ways (telecommunications,
the internet, computers, etc.), and its study has shed light on the
foundations of physics. Central to this is the idea that information
does not care how we choose to encode it -- we can encode information
on paper, in electronic pulses or carve it into stone. For almost all
of history our encoding of information has been into \emph{classical}
degrees of freedom.  However, Nature is quantum-mechanical and, in
recent years, we have begun to use \emph{quantum} degrees of freedom
to encode information. A central question therefore arises: does
information in quantum mechanics have the same properties as in
classical mechanics?

Now, the state of even the simplest quantum system -- a qubit -- is
specified by continuous parameters.  This means that it requires an
infinite amount of information to specify the state exactly.  For
example, the amplitude $\alpha$ of $|0\>$ in the superposition $\alpha
|0\> + \beta |1\>$ could encode the decimal expansion of $\pi$. Thus,
at first glance, it seems that that quantum systems can carry vastly
more information than classical systems.  However, Holevo
\cite{Holevo,Nielsen-Chuang, Wilde} showed only a single bit of
classical information can ever be \emph{extracted} from a qubit system
via measurement.  Further, in spite having a continuous infinity of
pure states, quantum computation do not suffer from the the problems
that rule out analog classical computers
\cite{Nielsen-Chuang}. Powerful theorems on the discretization of
errors \cite{Nielsen-Chuang} tell us that we do not need to correct a
continuum of errors, but only particular discrete types. These
surprising characteristics present a basic conundrum: how is it that
qubits behave as if they are discrete systems when their state space
forms a \emph{continuum}?

\begin{figure}[htb]
  \begin{center}
    \includegraphics[height=6cm]{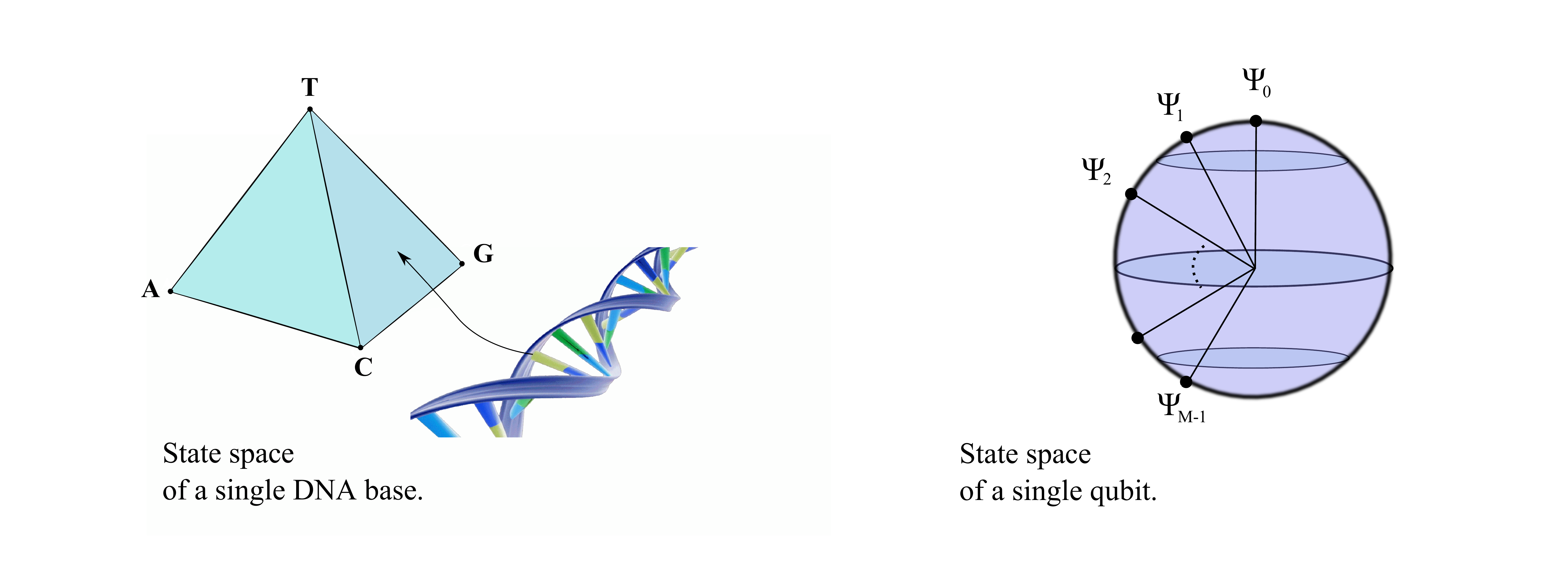}
  \end{center}
  \caption{\label{Hardystates}\textbf{The continuity of quantum states
      is non-classical:} In a noisy environment, a single DNA base has
    state space manifold given by a tetrahedron. Genetic information
    can be made robust against errors, despite having a continuous
    state space -- underlying discreteness exists.  Quantum
    information can also be made robust against errors, however unlike
    DNA, there can never be any underlying discreteness. Any
    statistical theory that reproduces the quantum predictions must
    have an infinity of microstates. Finite dimensional quantum
    systems necessarily contain an infinite amount of information, in
    contrast to classical systems with a finite state space.}
\end{figure}

As already discussed, in classical statistical mechanics we can
consider the allowed macrostates: the set of probability distributions
over some state space $\Lambda$ of microstates. It is easy to see that
these distributions \emph{also} form a continuum -- even if there is
only a discrete finite set of microstates. As an example, consider the
case of DNA bases, which can be in one of 4 microstates $A$, $T$, $C$
or $G$.  The macrostate for a single base is therefore a probability
distribution $\p = (p_A, p_T,p_C,p_G)$, obeying $\sum_j p_j =1$ and $0
\le p_j \le 1$ for all $j=A,T,C,G$. The set of such distributions
therefore forms a solid tetrahedron (a simplex) in 3-dimensional
space, and there is a continuum of macrostates (see
Figure~\ref{Hardystates}).

The fact that qubits behave in many ways like discrete, finite systems
would be easily explained if perhaps there were only a finite number
of more fundamental states -- like the finite number of DNA bases -- and if
the continuum of quantum states only represented our uncertainty about
which one of them is occupied -- like the continuum of DNA
macrostates.  Surprisingly, in spite of Holevo's bound and the
discretization of errors, this cannot be the case: any future physical
theory that reproduces the physics of \emph{finite}-dimensional
quantum systems must have an \emph{infinite} number of fundamental
states.

The proof, due to Hardy \cite{Hardy-baggage} (see
\cite{Massar-baggage} for an earlier related result), is quite
straightforward.  Firstly, for the sake of contradiction, assume that
$\Lambda$ is a finite set having $N$ elements, i.e.\ there are $N$
fundamental states in the theory.  Let $p(\lambda|\Psi)$ be the
probability distribution corresponding to $|\Psi\>$ in some
hypothetical future theory, and define the \emph{support} of
$p(\lambda|\Psi)$ to be
\begin{equation}
  \Lambda_{\Psi} = \{\lambda | p(\lambda|\Psi) > 0\}.
\end{equation}

Consider a measurement basis that includes the state $|\Psi\>$.  In
the underlying theory, the $|\Psi\>$ outcome is represented by a
conditional probability $p(\Psi|\lambda)$.  By the discrete version of
Eq.~\eqref{reproducegen}, with integrals replaced by sums, if we
prepare the state $|\Psi\>$ and measure in this basis, the underlying
theory must satisfy
\begin{equation}
  \label{certainreprod}
  \sum_{\lambda} p(\Psi|\lambda) p(\lambda|\Psi) = |\<\Psi|\Psi\>|^2 =
  1.
\end{equation}

Now note that $p(\lambda|\Psi)$, being a probability distribution,
must satisfy $\sum_{\lambda} p(\lambda|\Psi) = 1$. This means that
$p(\Psi|\lambda)$ must equal $1$ for all $\lambda \in \Lambda_{\Psi}$
in order to also make Eq.~\eqref{certainreprod} true.

Next, consider a two dimensional subspace spanned two orthonormal
states $|0\>$ and $|1\>$, and consider the $M$ states
\begin{equation}
  |\Psi_j \> = \cos \left ( \frac{j\pi}{2M} \right ) |0\> + \sin
  \left (\frac{j \pi}{2M} \right ) |1\>, 
\end{equation}
for $j = 0,1,2,\dots M-1$, as illustrated in Figure~\ref{Hardystates},
where $M$ can be chosen as large as we wish.  For any finite $M$,
these states satisfy,
\begin{equation}
  \label{overlap}
  |\<\Psi_k | \Psi_j\>|^2 < 1 
\end{equation}
for any $k \ne j$.

We will now show that reproducing the statistics of these states
implies that $\Lambda$ must contain at least $N = \log_2 M$ states and
hence, since $M$ can be chosen to be arbitrarily large, $N$ must be
infinite.  

Consider preparing the system in the state $|\Psi_j\>$ and measuring
it in a basis that contains $|\Psi_k\>$ for $k \neq j$.  Then,
Eq.~\eqref{overlap} implies that
\begin{equation}
  \sum_{\lambda} p(\Psi_k|\lambda)p(\lambda|\Psi_j) =
  |\<\Psi_k|\Psi_j\>| < 1. 
\end{equation}
This means that there must exist a $\lambda \in \Lambda_{\Psi_j}$ such
that $p(\Psi_k|\lambda) < 1$ since otherwise the sum would equal $1$.
Since $p(\Psi_k|\lambda) = 1$ everywhere on $\Lambda_{\Psi_k}$, this
means that $\Lambda_{\Psi_j}$ and $\Lambda_{\Psi_k}$ must be different
subsets of $\Lambda$.  Since this argument applies to every pair $k
\neq j$, we must in total have $M$ distinct subsets of $\Lambda$.

Now, if $\Lambda$ has $N$ elements then it has $2^N$ distinct subsets,
so we must have $2^N \geq M$, or $N \geq \log_2 M$.  However, $M$ can
be arbitrarily large, and since $\log_2 M \rightarrow \infty$ as $M
\rightarrow \infty$, we conclude that if the microstate description
reproduces quantum theory then $\Lambda$ must have infinitely many
elements. No finite set of states will ever work, and even the most
primitive quantum system must contain an infinite amount of
information -- in stark contrast with classical theory. One can
further prove that there must in fact be a continuous infinity of
microstates.  A heuristic way to see this is that the experimental
probability distributions for quantum states vary smoothly as we vary
the measurement basis, and so any underlying model must also inherit
this smoothness.  We refer the reader to the literature
\cite{Montina1, Montina2} for a rigorous proof of this.

It is said that discreteness is a distinctly quantum-mechanical
phenomenon, but anyone who has ever played a video game will tell you
that the concept of a discrete (pseudo-random) classical world is
really not so strange. Hardy's theorem together with results such as
Holevo's bound and the discreteness of errors, show that precisely the
opposite is the case: it is instead the \emph{continuity of quantum physics}
that is so strange. How can it be that we have a continuum of quantum
states that ostensibly behave discretely but we do not have, and
cannot have, an underlying discrete structure?

\subsection{The Pusey--Barrett--Rudolph Theorem: Is the wave-function $\Psi$ carved into a quantum system?}

\label{PBR}
 
In \S\ref{KS}, we alluded to a particularly subtle question: is the
wave-function an objective property of a quantum system? Again, our
meaning comes through comparison with classical statistical
mechanics. There the microstates are the objective properties of the
system -- at any instant of time the ``real'' state of the system is
actually a particular microstate, in contrast to the system's
macrostate which simply describes the ensemble properties of the
system and yields the thermodynamic variables of interest. Despite
there being no consensus on what, if any, objective properties exist
in the quantum realm, we can still ask whether it is possible for a
system to have the same objective properties when it is prepared in
one of two different quantum states. If the answer is ``no'', then we
know that the wave-function must be considered an intrinsic, or
objective property of a system.  Note that Hardy's theorem does not
settle the question of whether the wave-function is an intrinsic
property because, although the space of fundamental states must be a
continuum, the distributions corresponding to distinct quantum states
may still overlap, as they do in the Kochen-Specker model discussed in
\S\ref{KS}.
 
\begin{figure}[htb]
  \begin{center}
    \includegraphics[width=7.5cm]{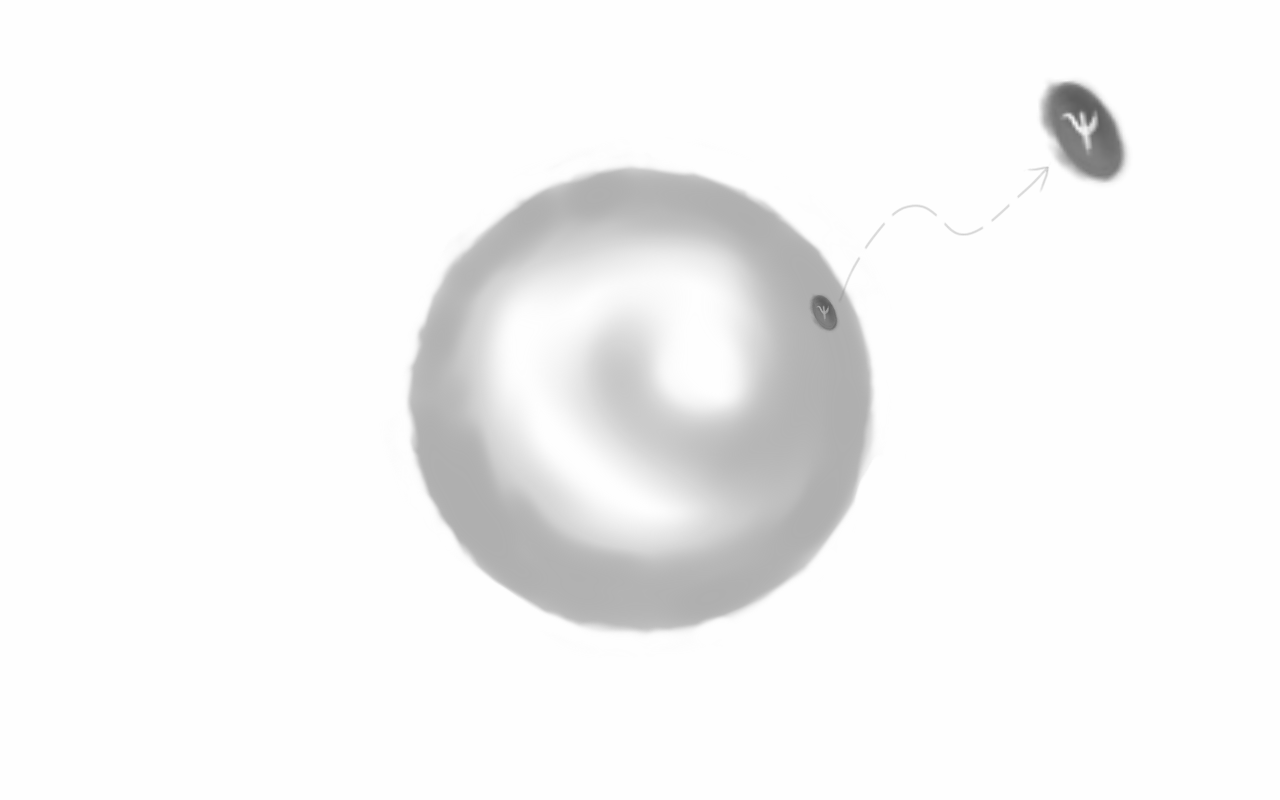}
    \caption{Is the wave-function $\Psi$ carved into a quantum system?}
  \end{center}
\end{figure}

The broad statistical framework allows us to frame this question in a
quantitative way. For any hypothetical theory with fundamental states
$\Lambda$ that reproduces the predictions of quantum mechanics for
some system, we say that the wave-function is an \emph{intrinsic}
property of the system if any two distinct quantum states $\Psi_1$ and
$\Psi_2$ have corresponding probability distributions
$p(\lambda|\Psi_1)$ and $p(\lambda|\Psi_2)$ that do not overlap. To
put it another way, consider the case where $\Lambda$ is a finite
set\footnote{Measure-theoretic qualifications are needed to deal with
  the general case.  See \cite{MattReview} for details.}.  Then, by
$p(\lambda|\Psi_1)$ and $p(\lambda|\Psi_2)$ overlapping, we mean that
there exists a microstate $\lambda_\star \in \Lambda$ for which both
$p (\lambda_\star |\Psi_1) > 0$ and $p (\lambda_\star|\Psi_2) > 0$.
If the system occupies $\lambda_*$ then we would not be able to tell
with certainty which of the two quantum states had been prepared, even
if we had access to the full microstate of the system.  Hence, it
would be impossible to ascribe a unique wave-function to this
microstate -- in such a hypothetical scenario the wave-function would
not be an intrinsic property of the system.

We now discuss a recent result due to Pusey, Barrett and Rudolph,
which shows that, under two additional reasonable assumptions, the
wave-function must be an intrinsic property of a quantum system
\cite{PBR}.

It suffices to consider a two dimensional quantum system, since, for
any higher-dimensional system, we can simply restrict to a
two-dimensional subspace. We establish a \emph{reducio ad absurdum} by
supposing there did exist some future statistical theory in which
$\Psi$ is a not an intrinsic property and from this arrive at a
contradiction.

Suppose that for two quantum states $|\Psi_1\>$ and $|\Psi_2\>$, the
corresponding distributions $p( \lambda |\Psi_1)$ and $p(\lambda
|\Psi_2)$ overlap. Again, we specialize to a finite space for
simplicity, so this means that there is an underlying microstate
$\lambda_\star$ which has a probability of occurring if we prepared
either the quantum state $|\Psi_1\>$ or the quantum state
$|\Psi_2\>$. More precisely, regardless of which of these states is
prepared, the microstate $\lambda_{\star}$ will be occupied a non-zero
fraction $P_\star >0$ of the time.

We now introduce the two assumptions used to prove the theorem.
Firstly, imagine preparing two copies of the system in one of the
\emph{four} quantum states
\begin{align}
  |\Psi_{11}\> & = |\Psi_1\> \otimes |\Psi_1\>, &
  |\Psi_{12}\> & = |\Psi_1\> \otimes |\Psi_2\> \nonumber \\ 
  |\Psi_{21}\> & = |\Psi_2\> \otimes |\Psi_1\> , &
  |\Psi_{22}\> & = |\Psi_2\> \otimes |\Psi_2\>. 
\end{align}
We can imagine that the two systems are initially located very far
apart, and Alice and Bob each choose whether to prepare $|\Psi_1\>$ or
$|\Psi_2\>$ independently of each other.

The first assumption is that, when two systems are prepared
independently like this, each system gets its own copy of $\Lambda$.
The total state space of the two systems is simply the product of two
copies of $\Lambda$ and so microstates are given by $(\lambda_1,
\lambda_2)\in \Lambda \times \Lambda$.  This means that the quantum
states $|\Psi_{jk}\>$ are associated with probability distributions of
the form $p(\lambda_1,\lambda_2|\Psi_{jk})$ and measurements on the
joint system are associated with conditional probability distributions
of the form $p(\Phi|\lambda_1,\lambda_2)$, where $|\Phi\>$ is a vector
in the basis we are measuring.

The second assumption is that the distribution describing the total
state $p (\lambda_1, \lambda_2|\Psi_{jk})$ factorizes as $p(\lambda_1,
\lambda_2 |\Psi_{jk}) = p(\lambda_1|\Psi_j) p (\lambda_2|\Psi_k)$,
where $p(\lambda_1|\Psi_j)$ and $p (\lambda_2|\Psi_k)$ are the
distributions that would be associated with $|\Psi_j\>$ and
$|\Psi_k\>$ for a single system. Taken together, these two assumptions
are called \emph{preparation independence}.

We now illustrate how preparation independence can be used to prove
that the quantum state must be an intrinsic property of a quantum
system by an example, before outlining the general result.  Suppose
that $|\Psi_1\> = |0\>$ and $|\Psi_2\> = |+\> = \frac{1}{\sqrt{2}}
\left ( |0\> + |1\> \right )$.  Under the assumption of preparation
independence, if we prepare one of the four states $|\Psi_{jk}\>$,
then, at least a fraction $P_\star^2 $ of the time, the joint system
will be in the microstate $(\lambda_{\star,1} , \lambda_{\star,2})$,
i.e.\ both systems will occupy the microstate $\lambda_{\star}$, for
which it is impossible to tell which of $|\Psi_1\>$ or $|\Psi_2\>$ was
prepared with certainty. This leads to the desired contradiction, once
we consider the following two-qubit measurement in the basis
consisting of the four vectors
\begin{align}
  |\Phi_{11}\> & = \frac{1}{\sqrt{2}} ( |0 \> \otimes |1\> + |1\> \otimes
  |0\>) \nonumber \\ 
  |\Phi_{12}\> & = \frac{1}{\sqrt{2}} ( |0 \> \otimes |-\> + |1\> \otimes
  |+\>) \nonumber \\ 
  |\Phi_{21}\> & = \frac{1}{\sqrt{2}} ( |+ \> \otimes |1\> + |-\> \otimes
  |0\>) \nonumber \\ 
  |\Phi_{22}\> & = \frac{1}{\sqrt{2}} ( |+ \> \otimes |-\> + |-\> \otimes
  |+\>),
\end{align}
where $|-\> = \frac{1}{\sqrt{2}} \left ( |0\> - |1\> \right )$.

Why is this particular (entangled) basis significant? The key point is
that $|\< \Phi_{jk}| \Psi_{jk}\>|^2 = 0$ for every choice of $j,k=1,2$. Put more
simply: the $|\Phi_{jk}\>$ measurement outcome never occurs when the
quantum state $|\Psi_{jk}\>$ is prepared. However, we know that,
whichever of these states is prepared, a fraction $P_\star^2$ of the
time the system is in the hypothetical microstate $(\lambda_{\star,1} ,
\lambda_{\star,2})$. If the system is in this microstate then which particular
outcome occurs when we make a measurement in this basis?  Suppose we
get the $|\Phi_{jk}\>$ outcome when the system occupies
$(\lambda_{\star,1} , \lambda_{\star,2})$.  We know that this
microstate occurs when $|\Psi_{jk}\>$ is prepared a non-zero fraction
of the time, so, in order to reproduce the quantum predictions, the
outcome $|\Phi_{jk}\>$ should \emph{never} occur for this
microstate. Thus, a non-zero fraction of the time the measurement
device cannot give any outcome that is consistent with quantum
mechanics, and we have our contradiction.

The above argument was specific to the states $|0\>$ and $|+\>$, but
it can be extended to show that every pair of pure states must
correspond to non-overlapping distributions.  Here, we will just
outline a version of this argument due to Moseley \cite{Moseley}, and
refer the reader to \cite{MattReview} for details.

First, note that if $|\<\Psi_1|\Psi_2\>|^2 \leq 1/2$, then the states
$|\Psi_1\>$ and $|\Psi_2\>$ are at least as distinguishable as $|0\>$
and $|+\>$, in the sense that it is at least as easy to tell them
apart via a quantum measurement.  It is known that, when this is the
case, it is possible to find a physical transformation that maps
$|\Psi_1\>$ to $|0\>$ and $|\Psi_2\>$ to $|+\>$ \cite{Chefles,
  Spehner}.  If we apply this transformation to both systems and then
make a measurement in the $|\Phi_{jk}\>$ basis then this whole
procedure can itself be thought of as a measurement on the states
$|\Psi_{jk}\>$.  This will have the same measurement probabilities as
before, so the previous argument can be adapted to this case.

It remains to deal with the case where $|\<\Psi_1|\Psi_2\>|^2 > 1/2$.
For this, we note that, if instead of preparing one system in the
state $|\Psi_1\>$ or $|\Psi_2\>$, we prepare $n$ systems either all in
the state $|\Psi_1\>$ or all in the state $|\Psi_2\>$, then the mod
squared inner product of the resulting states will be
$|\<\Psi_1|\Psi_2\>|^{2n}$.  For $|\<\Psi_1|\Psi_2\>|^2 < 1$, it is
possible to choose $n$ such that $|\<\Psi_1|\Psi_2\>|^{2n} < 1/2$ and
thus we can apply the previous argument to show that the distributions
corresponding to $|\Psi_1\>^{\otimes n}$ and $|\Psi_2\>^{\otimes n}$
can have no overlap.  However, preparation independence implies that
these distributions are just $n$-fold products of the distributions
corresponding to $|\Psi_1\>$ and $|\Psi_2\>$, so we infer that these
cannot have any overlap either.

The contradiction we have derived rules out a non-zero $P_\star$,
which quantifies how much the distributions corresponding to any pair
of pure states can overlap in \emph{any} future theory that can
reproduce the predictions of quantum theory. We conclude that there
can never exist a theory in which pure states have overlapping
distributions, unless preparation independence is violated
\cite{LJBR}. It is clear that preparation independence carries great
force, and is an extremely natural assumption to make -- imagine preparing a
Hydrogen atom close to its ground-state in one part of the universe,
and another person preparing another such atom on the other side of
the galaxy; no correlations ought to be created between these two
preparations, which should be fully independent. We have just shown
that this deceptively simple principle has a deep implication:
whatever future theory may arise, the quantum wave-function $\Psi$
will always be carved into a quantum system as an objective label of
reality.

\section{Conclusion}

\label{Conc}

To sum up, we have shown that many phenomena that are traditionally
viewed as intrinsically quantum-mechanical; such as randomness, discreteness, the
indistinguishability of states, measurement-uncertainty,
measurement-disturbance, complementarity, non-commutativity,
interference, the no-cloning theorem, and the collapse of the
wave-packet; all appear within classical statistical mechanics under
reversible dynamics.  These serve to map out classical fragments of
quantum physics, in a search for the genuinely strange aspects of the
theory. In addition to Bell's theorem on the failure of local
causality at a fundamental level, we have described two less
well-known results that reveal further deep and subtle insights into
the quantum realm.  Quantum systems unavoidably contain a continuous
infinity of information, despite their apparently discrete behaviour,
while the ability to prepare physical systems independently of one
another implies that the quantum wave-function is carved onto a quantum
system as an objective physical property of its microstate.

In this article, we have only presented a small sample of recent
results, in what is a flourishing area of current research.  For
example, we have not discussed the recently developed operational
approach to contextuality \cite{Spekkens-contextuality}, which is
another genuinely nonclassical quantum phenomenon, or the powerful
graph theoretic approach to contextuality \cite{CabelloGraph,
  AcinGraph}.  We have neglected many beautiful and deep results, such
as those of Colbeck and Renner \cite{Colbeck-Renner2, Colbeck-Renner3}
who rule out theories beyond quantum mechanics using weaker
assumptions than Bell; the results of Montina \cite{Montina-comm-comp,
  Montina-comm-comp2, Montina-comm-comp3, Montina-comm-comp4,
  Montina-comm-comp5} that reveal links between the structure of
classical fragments to the far more grounded topic of the
communication complexity of quantum channels; and, following on from
the Pusey--Barrett--Rudolph theorem, many other results and
experiments on the reality of the wave-function \cite{MattReview,
  Maroney1, Maroney2, Colbeck-Renner, Colbeck-Renner4, Leifer-Maroney,
  Aaronson2, Barrett, Leifer4, Branciard, PBR-exp, Colbeck-Renner-exp,
  Ringbauer, Mansfield, Allen}.  There are also ambitious programs
that seek to reformulate quantum theory as a theory of Bayesian
inference \cite{Leifer-Spekkens}, to derive quantum theory from
physically reasonable axioms \cite{HardyAxioms, HardyAxioms2, Dakic,
  Masanes, Chiribella}, and to formulate quantum theory in the absence
of fixed causal structure \cite{HardyCausaloid, HardyCausaloid2,
  Chiribella2, Oreshkov, Oreshkov2}.  Finally, we have not discussed
prominent areas of research such as quantum computing
\cite{Nielsen-Chuang}, quantum cryptography \cite{CryptoReview} and
quantum metrology \cite{MetrologyReview}, which are the practical
fruit of foundational investigations, and have their own insights to
offer about the difference between classical and quantum physics. For
these and more, we refer the interested reader to the bibliography,
where they will find an array of diverse and vibrant research programs
that continue to delve into the very foundations of quantum physics.

\section*{Acknowledgements}

DJ is supported by the Royal Society.  ML is supported by the
Foundational Questions Institute (FQXi).  Research at Perimeter
Institute is supported by the Government of Canada through Industry
Canada and by the Province of Ontario through the Ministry of Research
and Innovation.  DJ thanks Sania Jevtic for useful comments on an
earlier draft.

\appendix

\section{The Resolution Restriction on classical statistical
  mechanics}

\label{RR}

Any statistical Liouville distribution is a function $f$ on the
system's phase space (given by $\mathbb{R}^{2N}$ for a system of $N$
canonical coordinates), with $f(\x, \p) \ge 0$ and such that $\int
\!\! \d \x \d \p \, f( \x, \p) =1$. For simplicity we restrict
attention to a single particle in one dimension, but the same argument
holds for more general systems. The expectation value of the
particle's location is given by $\<x\> = \int \!\! \d x \d p \,
[f(x,p) x]$, while the expectation value of its momentum is $\<p\>:=
\int \!\!  \d x \d p \, [f(x,p) p]$.

The RR-constraint in phase space is imposed on the \emph{fluctuations}
about the mean $(\<x\>, \<p\>)$ for the Liouville distributions. To
this end, we define $z_1=x$ and $z_2 = p$, and a symmetric covariance
matrix $\gamma_{ij} := \< z_iz_j\> -\<z_i\> \<z_j\>$ that measures the
fluctuations of the Liouville distribution. In terms of position and
momentum it is given by
\begin{equation}
  \gamma = \begin{bmatrix} 
    (\Delta x)^2& \<x p\> - \<x\>\<p\> \\
    \<x p \> - \<x\> \<p\>& (\Delta p)^2
  \end{bmatrix},
\end{equation}
where $(\Delta x)^2 = \<x^2\> - \<x\>^2$ and $(\Delta p)^2 = \<p^2\>
- \<p\>^2$ are the variances of the position and momentum for the
Liouville distribution.

The RR-condition states that we only allow Liouville distributions
$f(x,p)$ that have some minimal level of fluctuations, as measured by
$\gamma$. We could impose this as a constraint on the eigenvalues of
$\gamma$, but a more elegant way is to demand that the matrix $\gamma$
obey the matrix equation
\begin{equation}
  \gamma + \alpha D \ge 0,
\end{equation}
for some $2\times 2$ matrix $D$, and some complex number $\alpha$ that measures the scale of the ``boxes'' on phase space,
and where, for a matrix $M$, $M \ge 0$ means that $M$ is a
semipositive matrix, i.e.~all the eigenvalues of $M$ are $\ge
0$. Since $\gamma$ is always semipositive, the case $\alpha=0$
corresponds to switching off the constraint.

Now, classical mechanics carries a symplectic structure
\cite{Goldstein}.  The dynamics generates a Liouville flow, which preserves this symplectic structure. However from the perspective of statistical mechanics, macroscopically reversible transformations correspond to the subset of linear symplectic transformations where the
canonical coordinates transform as $\boldsymbol{z} \rightarrow
A^\dagger \boldsymbol{z}$ for some symplectic matrix $A(t)$, obeying
$A^\dagger \Sigma A = \Sigma$.  The case of more general symplectic transformations is found to correspond to entropy production and irreversibility under the statistical restriction (see \cite{ERL-mech} and \cite{Goldstein} for more details). The matrix $\Sigma$ originates
from the classical Poisson bracket and is given by
\begin{equation}
  \Sigma = \begin{bmatrix}
    0 & -1 \\
    1 & 0 \\
  \end{bmatrix}.
\end{equation}
This is simply the symplectic matrix defining the group action of
reversible classical dynamics. Now, for the above RR-condition to be
meaningful it must be maintained under arbitrary dynamics. For the linear symplectic case the
matrix $\gamma$ transforms as $\gamma \rightarrow A(t)^\dagger \gamma
A(t)$, this implies that $A(t)^\dagger D A(t)$ should be equal to $D$
in order that the RR-condition has an invariant form under
dynamics. It is thus sufficient to choose $D=\Sigma$ within the
constraint. Moreover, $\Sigma$ is a skew-symmetric matrix with pure
imaginary eigenvalues and so, since $\gamma$ is a symmetric real
matrix, we must have that the constant $\alpha$ is pure imaginary in
order to obtain a meaningful constraint on the expectation values
(note the slightly different notation used here compared to the main
text, in which $i \lambda = \alpha$ and $C = iD$). We have argued for this identification under linear symplectic transformations, but it holds more generally. The
RR-condition therefore becomes
\begin{equation}
  \gamma + i \lambda \Sigma \ge 0,
\end{equation}
for some fixed minimal scale $\lambda \ge 0$ on the classical phase
space.

Finally we follow a statistical mechanics account of the physics
\cite{Tolman, Jaynes} so that, for a given covariance matrix $\gamma$,
we use the Gibbsian distribution $f$ that maximizes the thermodynamic
entropy $S = -\int\!\! \d x \d p \, f(x,p) \log f(x,p)$ and has
covariance matrix $\gamma$. Thus the scenario we have described is
precisely one of classical statistical mechanics where our classical
resolving power is bounded in phase space by a scale
$\lambda$. Indeed, for zero cross-correlations we find that $\Delta x
\Delta p \ge \lambda$, and so the RR-condition encodes a classical
uncertainty relation on the statistical system.

\section{General structure of quantum theory}

\label{QT}

In quantum theory, the state space of a physical system is a complex
Hilbert space $\H$.  A quantum state $|\Psi\>$ is a unit vector in
$\H$.  An observable is represented by a selfadjoint operator $A$,
$A^{\dagger} = A$.  Any such operator has a set of real eigenvalues
$a_j$, which represent the possible outcomes of a measurement of the
observable.  By the spectral theorem, $A$ can be written as $A =
\sum_j a_j P_j$, where $P_j$ is the projector onto the eigenspace
corresponding to $a_j$.  Note that, for an operator with continuous
spectrum, the sum would be replaced by an integral.  If each
eigenspace is one-dimensional, then $A$ is called non-degenerate and
the projectors can be written as $P_j = | \Phi_j\>\<\Phi_j|$, where
$|\Phi_j\>$ is the eigenvector corresponding to the eigenvalue $a_j$.
We shall only consider non-degenerate observables in what follows.

The eigenstates $|\Phi_j\>$ of a nondegenerate observable always form
a complete orthonormal basis for the Hilbert space $\H$, so a quantum
state $|\Psi\>$ can be decomposed in this basis as
\begin{equation}
  |\Psi\> = \sum_j \alpha_j |\Phi_j\>,
\end{equation}
where $\alpha_j = \<\Phi_j|\Psi\>$ and again the sum would be replaced
by an integral for an observable with a continuous spectrum.

If the system is prepared in the state $|\Psi\>$ and the observable
$A$ is measured, then the outcome $a_j$ is obtained with probability
$|\<\Phi_j|\Psi\>|^2 = |\alpha_j|^2$.  Assuming that the measurement
is performed in a non-destructive manner, after the measurement the
state is updated to $|\Phi_j\>$.  The transition from $|\Psi\>$ to
$|\Phi_j\>$ upon measurement is the notorious ``collapse of the
wave-packet''.

Since the measurement probabilities and the state-update rule only
depend on the eigenbasis $\{|\Phi_j\>\}$ of the observable and not on
the eigenvalues, they would remain the same if we measured a different
observable with the same eigenbasis.  For this reason, we shall often
speak of measuring the basis $\{|\Phi_j\>\}$, which simply means
measuring any non-degenerate observable that has this as its
eigenbasis.

The dynamics of a closed quantum system is described by the
Schr{\"o}dinger equation
\begin{equation}
  i \hbar \frac{\partial |\Psi(t)\>}{\partial t} = H |\Psi(t)\>, 
\end{equation}
where $H$ is the Hamiltonian, which is a self-adjoint operator.  If
the Hamiltonian is constant, then this leads to the formal solution
$|\Psi(t)\> = U(t-t_0)|\Psi(t_0)\>$, where $U(t) = e^{-i H t/\hbar}$.
It is easy to check that $U(t)$ is unitary, which means that
$U^{\dagger}(t)U(t) = U(t)U^{\dagger}(t) = \I$, where $\I$ is the
identity operator on $\H$.  Indeed, in principle any unitary operator
can be obtained by an appropriate choice of $H$.  If the Hamiltonain
varies in time, then the dynamics is still given by a unitary operator
$U(t)$, which is now given in terms of time-ordered exponentials of
the Hamiltonian, but the key point is that it is still unitary.
Because of these facts, we can simply say that the discrete-time
dynamics of a closed quantum system is described by a unitary operator
$U$.

Finally, if a quantum system consists of two subsystems with Hilbert
spaces $\H_1$ and $\H_2$, then the composite system has a Hilbert
space $\H_1 \otimes \H_2$.  This means that, if $\{|\Psi_j\>\}$ is a
basis for $\H_1$ and $\{|\Phi_j\>\}$ is a basis for $\H_2$, then $\H_1
\otimes \H_2$ is the Hilbert space spanned by the basis $\{|\Psi_j\>
\otimes |\Phi_k\>\}$.

\section{Qubits}

\label{Qubits}

A qubit is any quantum system with a two dimensional Hilbert space.
For example, it may be the spin of a spin-$1/2$ particle, the
polarization of a photon, or the energy levels of a two-level atom.
We use the notation $|0\> = (1,0)^T$, $|1\> = (0,1)^T$ to represent
the two standard basis states, which are the eigenstates of the Pauli
operator
\begin{equation}
  \sigma_3 = \left ( \begin{array}{rr} 1 & 0 \\ 0 & -1 \end{array}
  \right ). 
\end{equation}
For a spin-$1/2$ particle, these would be the spin-up and spin-down
states in the $z$ direction, and the notation $|z \uparrow\> = |0\>$,
$|z \downarrow\> = |1\>$ is often used in this case, but bear in mind
that the physical interpretation of these states depends on which
instantiation of a qubit we are considering.

The two other Pauli operators are
\begin{align}
  \sigma_1 & = \left ( \begin{array}{rr} 0 & 1 \\ 1 & 0 \end{array}
  \right ), &
  \sigma_2 & = \left ( \begin{array}{rr} 0 & -i \\ i & 0 \end{array}
  \right ),
\end{align}
which, for a spin-$1/2$ particle, represent spin in the $x$ and $y$
directions respectively.  The eigenstates of $\sigma_1$ are $| \pm \>
= \frac{1}{\sqrt{2}} \left ( |0\> \pm |1\>\right )$ and the
eigenstates of $\sigma_2$ are $| \pm i \> = \frac{1}{\sqrt{2}} \left (
  |0\> \pm i |1\> \right )$.  In the spin-$1/2$ instantiation, these
are often alternatively written as $|x \uparrow\> = |+\>$, $|x
\downarrow\> = |-\>$, $|y \uparrow\> = |+i\>$, and $|y \downarrow\> =
|-i\>$.

The Pauli operators are paradigmatic examples of noncommuting
operators.  Their commutators are given by $[\sigma_1,\sigma_2] = i
\sigma_3$, $[\sigma_2,\sigma_1] = -i\sigma_3$ and cyclic permutations.
However, noncommutativity is just a mathematical statement, so the
following facts are often pointed to as physical correlates of
noncommutativity.

\begin{figure}
\begin{center}
\includegraphics[width=5cm]{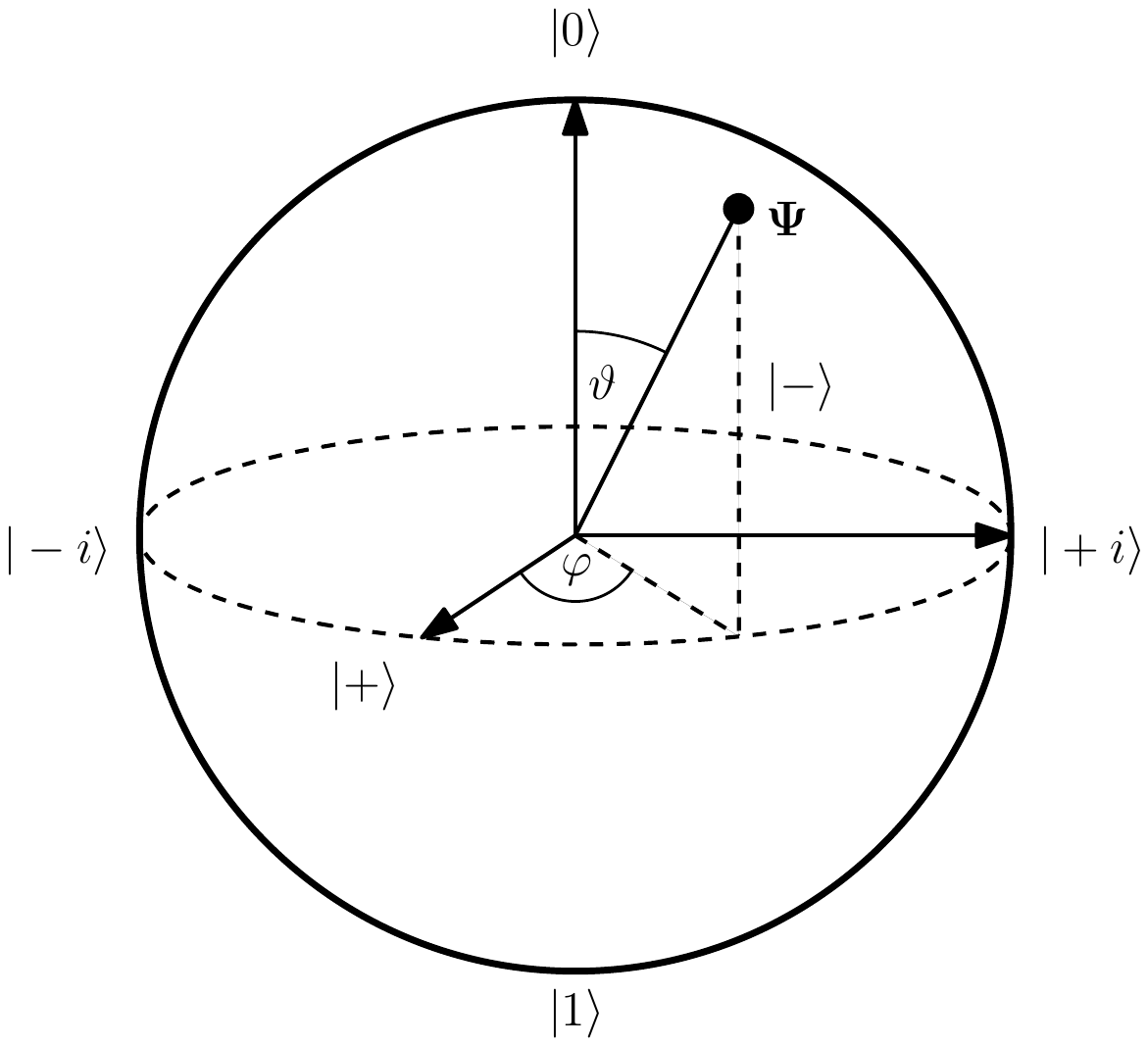}
\includegraphics[width=7cm]{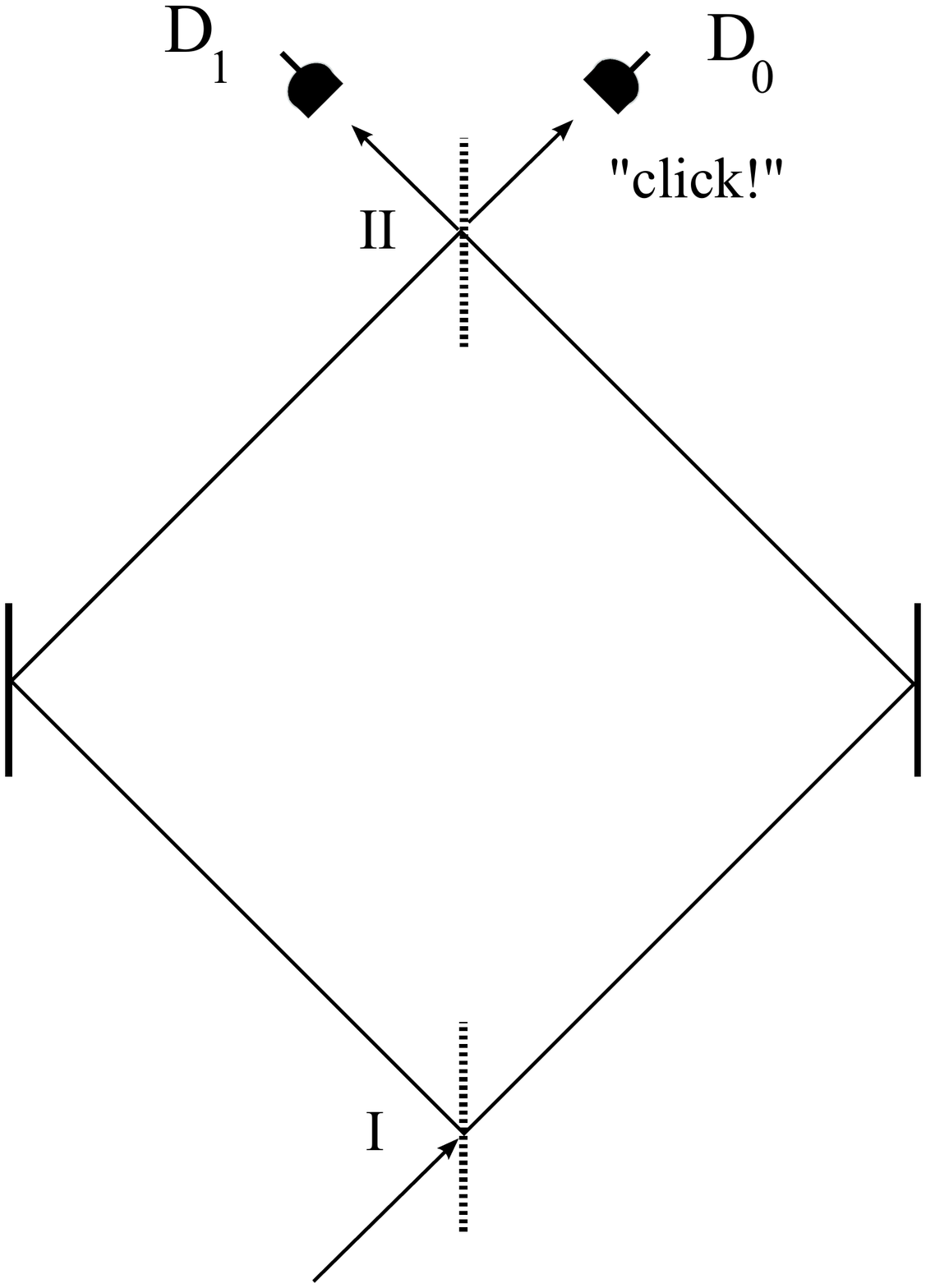}
\end{center}
\vspace{1cm}
\caption{\label{Bloch-Mach}\textbf{Qubit Systems:} The Bloch Sphere
  (left) is the state
  space of the most elementary quantum system -- the qubit. Every
  point in the sphere corresponds to a unique quantum state. The
  Mach-Zehnder interferometer (right) illustrates the process of
  interference in quantum mechanics. An in-coming photon strikes a
  50/50 beam-splitter I and splits into two different paths that later
  recombine (after mirror reflections) at a second beam splitter II.}
\end{figure}
\begin{itemize}
\item The Pauli observables are complementary to one another: If the
  system is in a state where $\sigma_3$ has a definite value, i.e.\
  either $|0\>$ or $|1\>$, then measurements in the bases
  $\{|+\>,|-\>\}$ and $\{|+i\>, |-i\>\}$ are completely uncertain in
  the sense that each outcome has probability $1/2$, and the same
  property holds under permutation of the Pauli operators.
\item If several Pauli measurements are performed in a row then the
  outcome depends on the order in which the observables are measured.
  For example, if the system is prepared in the state $|+\>$ and
  immediately measured in the basis $\{|+\>, |-\>\}$ then the $|+\>$
  outcome will be obtained with certainty.  However, if a measurement
  in the $\{|0\>,|1\>\}$ basis is performed first then each outcome of
  a subsequent measurement in the $\{|+\>,|-\>\}$ basis occurs with
  probability $1/2$.  To see this, suppose that the $|0\>$ outcome is
  obtained in the first measurement.  Then, the state gets updated to
  $|0\>$, which now has a probability $1/2$ of giving either outcome
  in a measurement in the $\{|+\>,|-\>\}$ basis.  The same is true if
  the first measurement yields the $|1\>$ outcome.  This phenomenon is
  often referred to by the phrase ``measurements disturb the state of
  the system'', but it is better to simply say that the order of
  measurements can affect the outcome statistics.
\end{itemize}

Another paradigmatic quantum phenomenon is interference, and this can
also be exhibited with the Pauli eigenstates.  Consider the
Mach-Zehnder interferometer with $50/50$ beam-splitters illustrated in
Figure~\ref{Bloch-Mach} and suppose that a single photon is passed through
it.  The two input ports can be represented on a two-dimensional
Hilbert space, where $|0\>$ represents a photon incident on the
beam-splitter from the left side and $|1\>$ from the right.  After the
beam-splitter, we use the same labels for the transmitted beams and
the opposite for reflected beams, so $|0\>$ is used to represent a
photon travelling along the right arm of the interferometer and $|1\>$
for the left, and after the second beamsplitter, $|0\>$ represents a
photon in the left output beam and $|1\>$ the right.  With these
conventions, the transformation implemented by a beam-splitter is
represented by the unitary matrix
\begin{equation}
  \frac{1}{\sqrt{2}} \left ( \begin{array}{rr} 1 & 1 \\ 1 &
      -1 \end{array} \right ),
\end{equation}
which maps $|0\>$ to $|+\> = \frac{1}{\sqrt{2}}(|0\> + |1\>)$ and
$|1\>$ to $|-\> = \frac{1}{\sqrt{2}} (|0\> - |1\>)$, and the detectors
perform a measurement in the $\{|0\>,|1\>\}$ basis, where $D_j$ firing
corresponds to the $|j\>$ outcome.

First consider what happens when we remove the second beam-splitter
and input the photon from the left of the first beam-splitter.  Then,
$|0\>$ gets mapped to $|+\>$ and there is a probability $1/2$ that
each of the detectors will fire.  These statistics are consistent with
the idea that, at each beam-splitter, the photon is either definitely
transmitted or definitely reflected, with probability $1/2$ each.
However, if we replace the second beam-splitter, then $|+\>$ gets
mapped to $|0\>$ before the detection and so $D_0$ will fire with
certainty.  In other words, there is constructive interference between
the two beams at the left output port of the interferometer, and
destructive interference on the right.  This is not consistent with
the idea that, at a beamsplitter, the photon always goes one way or
the other with probability $1/2$, since otherwise we would expect both
detectors to fire with probability $1/2$ in this case as well.

Finally, we also need to consider how a general qubit state is
represented.  In the $\{|0\>,|1\>\}$ basis, an arbitrary state of a
qubit can be written as
\begin{equation}
  |\Psi\> = \cos \left ( \frac{\vartheta}{2} \right ) |0\> + e^{i\varphi}
  \sin \left ( \frac{\vartheta}{2} \right ) |1\>,
\end{equation}
where $0 \leq \vartheta \leq \pi$ and $-\pi \leq \varphi \leq \pi$.
Alternatively, because the Pauli operators span the vector space of
$2\times 2$ matrices, we can represent the state by its \emph{Bloch
  vector} $\boldsymbol{\Psi} = (\<\Psi|\sigma_1|\Psi\>,
\<\Psi|\sigma_2|\Psi\>, \<\Psi|\sigma_3|\Psi\>) = (\sin \vartheta \cos
\varphi, \sin \vartheta \sin \varphi, \cos \vartheta)$.  This shows
that the states of a qubit are isomorphic to points on the surface of
the unit sphere $S^2$ in real three-dimensional space, which is known
in this context as the \emph{Bloch sphere} (see
Figure~\ref{Bloch-Mach}).

Note that, due to the doubling of the angles in the Bloch sphere
representation, states that are orthogonal in the Hilbert space are
represented by pairs of antipodal points on the sphere.  This means
that an orthonormal basis $\{|\Phi\>, |\Phi^{\perp}\>\}$,
representing a measurement, is represented by a pair
$\{\boldsymbol{\Phi}, \boldsymbol{\Phi}^{\perp}\}$ of antipodal points
on the sphere.  The probabilities that quantum theory predicts for
measurement outcomes can also be rewritten in terms of Bloch vectors
via   $|\<\Phi|\Psi\>|^2 = \frac{1}{2} \left ( 1 + \boldsymbol{\Phi} \cdot
    \boldsymbol{\Psi} \right )$.

\section{Continuous variable systems}

\label{CVS}

In nonrelativistic quantum mechanics, the Hilbert space of a spinless
particle in one dimension spanned by the continuum of eigenstates
$|x\>$ of the position operator $\hat{x}$, $\hat{x} |x\> = x|x\>$.  A
state $|\Psi\>$ is therefore written as
\begin{equation}
  | \Psi \> = \int_{-\infty}^{\infty}\!\! \d x \, \psi(x) | x \>,
\end{equation}  
where $\psi(x) = \<x|\Psi\>$ is the wave-function in position
representation and $|\psi(x)|^2\d x$ is the probability that the outcome
of a position measurement will be between $x$ and $x + \d x$. 

The momentum operator $\hat{p}$ is represented as $\hat{p} = -i\hbar
\frac{\partial}{\partial x}$ and, using this, we can alternatively
represent $|\Psi\>$ in the basis of momentum eigenstates as
\begin{equation}
  | \Psi \> = \int_{-\infty}^{\infty}\!\! \d p \, \phi(p) | p \>,
\end{equation} 
where $|\phi(p)|^2dp$ is the probability that the outcome of a
momentum measurement will be between $p$ and $p + dp$.  The functions
$\psi(x)$ and $\phi(p)$ are related by the Fourier transform.

Position and momentum are complementary variables, in the sense that
if one of them is certain then the other is completely indeterminate.
Further, they obey the Heisenberg uncertainty relation
\begin{equation}
  \Delta x \Delta p \geq \frac{\hbar}{2},
\end{equation}
where $\Delta x = \sqrt{\<x^2\> - \<x\>^2}$ and $\Delta p =
\sqrt{\<p^2\> - \<p\>^2}$ are the standard deviations of position and
momentum computed for any fixed state $|\Psi\>$.

Although it may be less familiar, it is possible to represent a
quantum state as a function $W(x,p)$ on phase space, known as the
Wigner function, defined by
\begin{equation}
  W(x,p) = \frac{1}{2\pi\hbar} \int_{-\infty}^{\infty}\! \! \d y \, \psi^{*}\left (
    x + \frac{y}{2} \right ) \psi \left ( x - \frac{y}{2} \right )
  e^{-\frac{ipy}{\hbar}}. 
\end{equation}
The Wigner function is like a probability density on phase space,
except that it can take negative values.  Nevertheless, its marginals
give the correct probability densities over position and momentum,
i.e.\ 
\begin{align}
  W(x) & = \int_{-\infty}^{\infty}\!\!  \d p \, W(x,p) = |\psi(x)|^2 & W(p) & =
  \int_{-\infty}^{\infty}\!\! \d x \,W(x,p) = |\phi(p)|^2.
\end{align}

A \emph{Gaussian quantum state} is a state for which the Wigner
function is a Gaussian.  Gaussian states are an important subclass of
quantum states that occur in a variety of applications, e.g.\ the
coherent states produced by a laser are Gaussian.  In order to write
down a Gaussian state it is helpful to form the phase space vector
$\boldsymbol{z} = (x,p)^T$ and define the $2\times 2$ covariance
matrix $\gamma$ with elements $\gamma_{jk} = \<z_jz_k\> - \<z_j\>\<z_k\>$, i.e.\ 
\begin{equation}
  \gamma = \left ( \begin{array}{cc}(\Delta x)^2 & \<xp\>-\<x\>\<p\> \\ \<xp\> -\<x\> \<p\> &
     (\Delta p)^2 \end{array} \right ),
\end{equation}
where $\Delta x$ and $\Delta p$ are the standard deviations defined above.

Then, a Gaussian state has a Wigner function of the form
\begin{equation}
  W(x,p) \propto e^{-\frac{1}{2} (\boldsymbol{z} -
    \<\boldsymbol{z}\>)^T \gamma^{-1}(\boldsymbol{z} -
    \<\boldsymbol{z}\>)}, 
\end{equation}
where $\<\boldsymbol{z}\> = (\<x\>,\<p\>)^T$.

In general, a continuous variable system with $N$ degrees of freedom
is represented by a Hilbert space spanned by vectors
$|\boldsymbol{x}\>$, where $\boldsymbol{x} \in \mathbbm{R}^N$ is a
configuration space point.  A state $|\Psi\>$ is of the form
\begin{equation}
  | \Psi \> = \int \d \boldsymbol{x} \, \psi(\boldsymbol{x}) |
  \boldsymbol{x} \>, 
\end{equation}  
where the wave-function $\psi(x) = \<\boldsymbol{x}|\Psi\>$ is now a
function on configuration space and $|\psi(\boldsymbol{x})|^2$ is the
probability density of finding the system in a given configuration.

We can again define a Wigner function as
\begin{equation}
  W(\boldsymbol{x},\boldsymbol{p}) = \frac{1}{(2\pi\hbar)^N} \int
  \d \boldsymbol{y} \, \psi^{*}\left ( 
    \boldsymbol{x} + \frac{\boldsymbol{y}}{2} \right ) \psi \left (
    \boldsymbol{x} - \frac{\boldsymbol{y}}{2} \right ) 
  e^{-\frac{i\boldsymbol{p} \cdot \boldsymbol{y}}{\hbar}},
\end{equation}
where the integration is now over configuration space and $d$ is the
number of spatial dimensions.  Defining the vector $\boldsymbol{z} =
(x_1,p_1,x_2,p_2,\ldots,x_N,p_N)$ and covariance matrix $\gamma_{jk} =
\<z_jz_k\> -\<z_j\>\<z_k\>$, a Gaussian state is again one that has a Wigner function
of the form
\begin{equation}
  W(\boldsymbol{x},\boldsymbol{p}) \propto e^{-\frac{1}{2}
    (\boldsymbol{z} - \<\boldsymbol{z}\>)^T \gamma^{-1}(\boldsymbol{z} -
    \<\boldsymbol{z}\>)}. 
\end{equation}

Finally, there is also a notion of Gaussian measurements and Gaussian
dynamics.  A Gaussian measurement is one that, when performed on a
Gaussian state, the state remains Gaussian after the measurement, and
a Gaussian unitary operator is a unitary that maps Gaussian states to
Gaussian states.  Importantly, this is also required to hold when the
operation is only applied to a subsystem.  For example, when applied
to a two-particle Gaussian state, performing a one-particle Gaussian
transformation to one of the particles and doing nothing to the other
should leave the system in a two-particle Gaussian state.  By
\emph{Gaussian Quantum Mechanics}, we mean the sub-theory of continuous
variable quantum mechanics in which states, measurements, and dynamics
are all restricted to be Gaussian.

\end{document}